%% file: main.tex
\documentclass[sigconf,authorversion,screen]{acmart}

\AtBeginDocument{%
  \providecommand\BibTeX{{%
    \normalfont B\kern-0.5em{\scshape i\kern-0.25em b}\kern-0.8em\TeX}}}

\copyrightyear{2022}
\acmYear{2022}
\setcopyright{rightsretained}
\acmConference[CHI '22]{CHI Conference on Human Factors in Computing Systems}{April 29-May 5, 2022}{New Orleans, LA, USA}
\acmBooktitle{CHI Conference on Human Factors in Computing Systems (CHI '22), April 29-May 5, 2022, New Orleans, LA, USA}
\acmDOI{10.1145/3491102.3517497}
\acmISBN{978-1-4503-9157-3/22/04}

\newcommand{\qt}[1]{\textit{``#1''}}
\newcommand{\qtp}[2]{\textit{``#1''} (#2)}
\newcommand{\dataseturl}{~\url{https://github.com/google-research/google-research/tree/master/taperception}}

\usepackage{changes}
\begin{document}

\title{Predicting and Explaining Mobile UI Tappability with Vision Modeling and Saliency Analysis}

\author{Eldon Schoop}
\authornote{This work was completed while the author was an intern at Google.}
\email{eschoop@berkeley.edu}
% \orcid{XXXX}
\affiliation{%
  \institution{UC Berkeley EECS}
  \city{Berkeley}
  \state{CA}
  \country{USA}
}

\author{Xin Zhou}
\email{zhouxin@google.com}
\affiliation{%
  \institution{Google Research}
  \city{Mountain View}
  \state{CA}
  \country{USA}
}

\author{Gang Li}
\email{leebird@google.com}
\affiliation{%
  \institution{Google Research}
  \city{Mountain View}
  \state{CA}
  \country{USA}
}

\author{Zhourong Chen}
\email{czhrong@gmail.com}
\affiliation{%
  \institution{Google Research}
  \city{Mountain View}
  \state{CA}
  \country{USA}
}

\author{Bj\"orn Hartmann}
\email{bjoern@eecs.berkeley.edu}
\affiliation{%
  \institution{UC Berkeley EECS}
  \city{Berkeley}
  \state{CA}
  \country{USA}
}

\author{Yang Li}
\email{yangli@acm.org}
\affiliation{%
  \institution{Google Research}
  \city{Mountain View}
  \state{CA}
  \country{USA}
}

% \renewcommand{\shortauthors}{Schoop, Xhou, Li, et al.}

\begin{abstract}
    UI designers often correct false affordances and improve the discoverability of features when users have trouble determining if elements are tappable. We contribute a novel system that models the perceived tappability of mobile UI elements with a vision-based deep neural network and helps provide design insights with dataset-level and instance-level explanations of model predictions. Our system retrieves designs from similar mobile UI examples from our dataset using the latent space of our model. We also contribute a novel use of an interpretability algorithm, XRAI, to generate a heatmap of UI elements that contribute to a given tappability prediction. Through several examples, we show how our system can help automate elements of UI usability analysis and provide insights for designers to iterate their designs. In addition, we share findings from an exploratory evaluation with professional designers to learn how AI-based tools can aid UI design and evaluation for tappability issues.
    % UI designers often iterate designs to correct false affordances and improve the discoverability of features when users have trouble determining if elements are tappable. We contribute a novel, automated system that models the perceived tappability of mobile UI elements with a pure vision-based deep neural network and provides design insights with dataset-level and instance-level explanations of model predictions.
    % Beyond predicting the tappability of a given element in a UI screenshot, our system retrieves designs from similar mobile UI examples from our dataset using the latent space of our model. We also contribute a novel use of a region-based saliency method to generate a heatmap of UI elements that contribute to the model's tappability prediction for a given input. Through several examples, we show how our system can help automate elements of UI usability analysis and provide design insights for designers to iterate their designs. \added{In addition, we conduct an exploratory evaluation of our system with 13 professional UI/UX designers to learn how our system may be used in UI design practice, which generated initial insights into how such an AI-based tool can aid UI design and evaluation for tappability issues.}
\end{abstract}

\begin{CCSXML}
<ccs2012>
<concept>
<concept_id>10003120.10003138.10003142</concept_id>
<concept_desc>Human-centered computing~Ubiquitous and mobile computing design and evaluation methods</concept_desc>
<concept_significance>500</concept_significance>
</concept>
<concept>
<concept_id>10003120.10003121.10003122</concept_id>
<concept_desc>Human-centered computing~HCI design and evaluation methods</concept_desc>
<concept_significance>500</concept_significance>
</concept>
</ccs2012>
\end{CCSXML}

\ccsdesc[500]{Human-centered computing~Ubiquitous and mobile computing design and evaluation methods}
\ccsdesc[500]{Human-centered computing~HCI design and evaluation methods}

\keywords{Mobile UIs, Deep Learning, Explainable AI, Interpretability}

\begin{teaserfigure}
    \centering
    \includegraphics[width=0.7\textwidth]{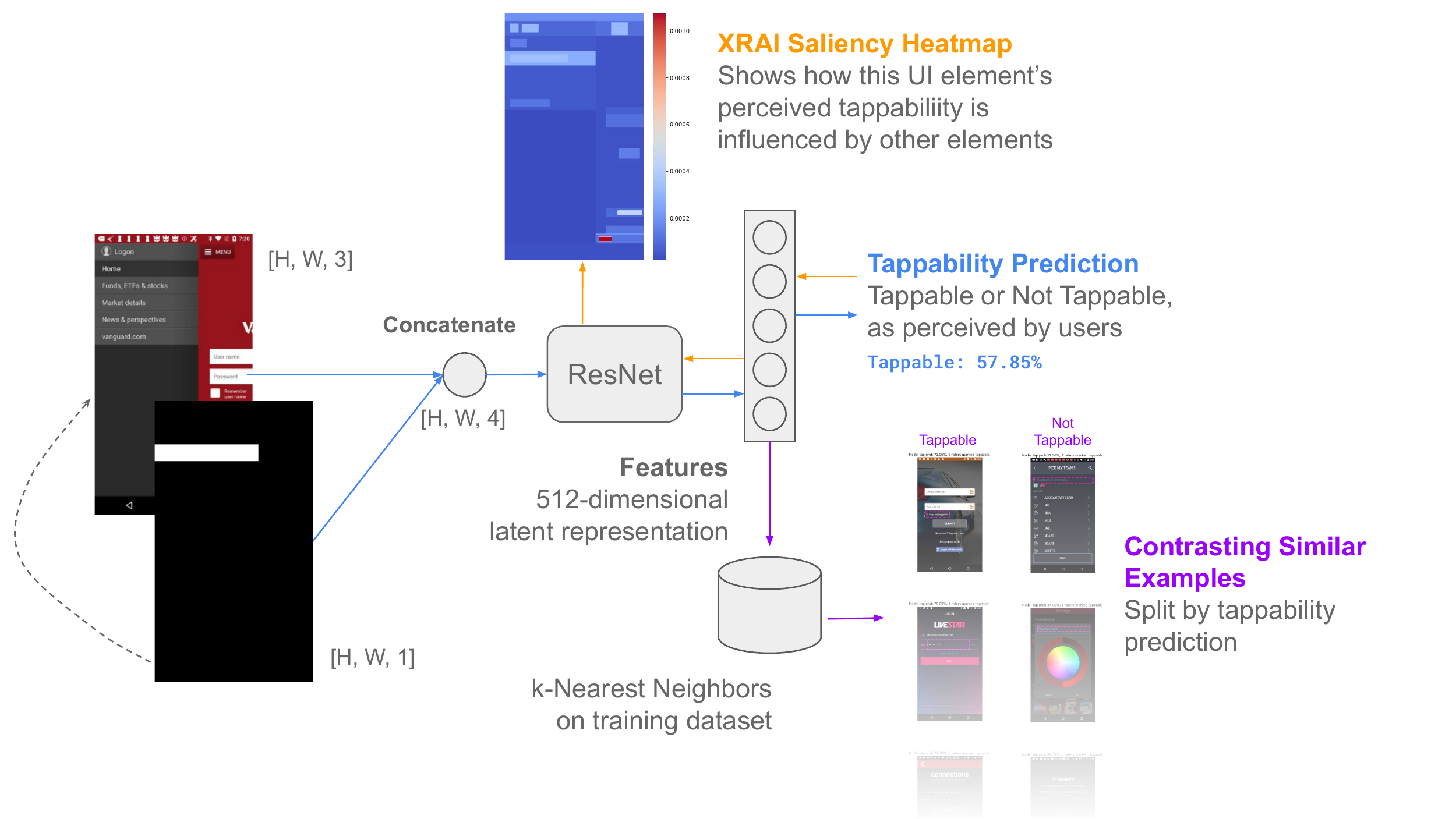}
    \caption{We use a deep learning based approach to predict whether a selected element in a mobile UI screenshot will be perceived by users as tappable, based on pixels only instead of view hierarchies required by previous work.
    To help designers better understand model predictions and to provide more actionable design feedback than predictions alone, we additionally use ML interpretability techniques to help explain the output of our model. We use XRAI to highlight areas in the input screenshot that most strongly influence the tappability prediction for the selected region, and use k-Nearest Neighbors to present the most similar mobile UIs from the dataset with opposing influences on tappability perception.}
%   \Description{todo}
  \label{fig:teaser}
\end{teaserfigure}

\maketitle

\input{intro}
\input{relatedwork}
\input{dataset}
\input{model}
\input{explanations}
\input{analysis}
\input{eval}
\input{discussion}
\input{conclusion}

% \begin{acks}
% \end{acks}

\bibliographystyle{ACM-Reference-Format}
\bibliography{cites}

\newpage
\onecolumn
\appendix

\input{study_UIs}

\end{document}

%% file: intro.tex
\section{Introduction}

Tapping is a fundamental gesture in mobile User Interfaces (UIs). However, because of the highly varied styles of mobile UIs, users can have difficulty telling if UI elements are tappable~\cite{swearngin_modeling_2019}. This harms the usability of applications, e.g., when false affordances suggest an item is tappable when it is not; or when the design of a new feature limits its discoverability.

UI designers and User Experience (UX) researchers traditionally run user studies to evaluate the usability of their designs. While these studies can provide actionable feedback and lead to significant design insights, they are often costly and time-consuming to conduct.
% HCI research has developed several automated techniques to help designers more quickly evaluate their applications.
Recent works have applied Deep Learning (DL) techniques to predict whether users will correctly estimate if mobile UI elements are tappable~\cite{swearngin_modeling_2019} and predict user engagement with mobile UI animations~\cite{wu_animations_2020}.
These automated approaches can help designers gain quick insights into the usability of their applications, but lack the design guidance and explanations that can be gained from controlled user studies.
In addition, many automated tools rely on a functional mobile application or UIs with detailed specifications, such as view hierarchies, meaning that they may not be able to produce usable results on mockups. Yet, gaining feedback in the early stages of design is crucial.

The goal of this work is to produce a model that faithfully approximates the perception of real users for rapid, automated tappability evaluations, and a system which provides explanations of its predictions that offer insight for improving designs.
To gain a basis for understanding tappability perception at scale, we create a new dataset of crowdworkers' estimates of the tappability of UI elements in thousands of mobile UI screenshots from the RICO dataset~\cite{deka_rico_2017}. As shown in previous work~\cite{swearngin_modeling_2019}, human perceptions of tappability can vary significantly. To account for this, our new dataset includes 5 crowdworkers' labels for each UI element, by which we can more reliably estimate user perceptions at scale.
We use this dataset to train a purely vision-based deep neural network that, given a screenshot and a selected region of interest, predicts the perceived tappability of the selected UI element. This allows designers to rapidly assess how users may perceive elements of a mobile UI design, whether or not it is implemented in an application.

We take an important step further beyond tappability prediction by drawing upon techniques in Machine Learning (ML) interpretability and Explainable Artificial Intelligence (XAI) to explain our model's predictions in two ways~\cite{local_global}.
We provide a \textit{local} explanation which highlights regions in the input screenshot, indicating areas the model considers most important to the tappability prediction for a given element.
We provide a \textit{global} explanation which uses the latent space of our model to find contrasting nearest-neighbor examples in our source dataset, allowing users to discover patterns in visually similar UIs that have opposing influences on tappability perception.
To evaluate our model and its explanation outputs, we share an in-depth analysis of the behavior of our model using random examples from our source dataset, and conduct an exploratory evaluation to seek feedback from professional UI/UX designers.

Specifically, this paper contributes:
\begin{itemize}
    \item A new dataset collecting tappability labels from multiple crowdworkers per example on thousands of mobile application screenshots\footnote{We release our dataset publicly at \dataseturl.}. This extends previous work~\cite{swearngin_modeling_2019} to better address human uncertainty in tappability perception;
    \item A vision-based deep neural network that predicts the perceived tappability of selected UI element(s) in a mobile UI screenshot by only relying on pixels. Our model is capable of examining UI designs that are not fully specified (e.g., mockups).
    This significantly extends prior work since it enables a broader set of applications, e.g., to produce feedback for early-stage designs;
    \item A novel method for eliciting explanations of tappability predictions from our model by annotating the screen under inspection, and by surfacing similar examples from the dataset that have opposing influences on tappability perception;
    \item An in-depth analysis of model behavior on randomly selected examples from an evaluation dataset, and an exploratory evaluation with 13 professional UI/UX designers, from which we distill initial insights into how an AI-based tool can assist designers.
\end{itemize}

%% file: relatedwork.tex
\section{Related Work}

Our work builds on three primary areas: automated tools which assist UI designers in exploring and evaluating UIs; automated tools which assist in evaluating the usability of UIs; and algorithms and methods for interpreting the predictions of deep neural networks.

\subsection{Data-Driven UI Design and Exploration}
The HCI community has produced many research artifacts that help designers create UIs through the collection and use of large-scale UI datasets~\cite{ai_hci_workshop}.
Datasets such as ERICA~\cite{erica} and RICO~\cite{deka_rico_2017} have enabled the creation of numerous data-driven systems in this domain.
While the vast size of RICO has made it attractive for data-driven applications in research, it is known to have significant label noise~\cite{li_grounding}. Many works add annotations to RICO or take additional cleaning steps, e.g., ENRICO, which organizes RICO into design topics~\cite{enrico}, and RICO\textsubscript{clean} which relabels icon elements in the original dataset~\cite{zang_icons}.
Our work contributes a dataset that augments a cleaned subset of RICO with annotations from multiple crowdworkers predicting the tappability of various UI elements.

Designers benefit from viewing selections of varied UI design examples to serve as inspiration in the design process~\cite{amanda_scout}. Gallery DC uses a neural network to tag elements in mobile UI screenshots, presenting them in a gallery to help designers explore a large set of UI element examples~\cite{chen_gallery_2019}.
Other works help designers retrieve examples from datasets like RICO, e.g., from hand-drawn sketches~\cite{swire}, low-fidelity wireframes~\cite{chen_wireframe-based_2020}, and text-annotated layout information~\cite{screen2vec, UIBert, ActionBERT}.
We also use the latent space of a deep neural network for UI retrieval. However, our model is trained on the perception of human raters, rather than to reconstruct UI layouts. This means that retrieved examples are similar in how they are perceived by humans to be tappable, rather than in visual similarity alone. In addition, our model uses the raw pixels of a mobile UI as input, allowing it to capture more detailed visual features than layouts.

\subsection{Computationally Mediated UI Evaluation}
Because of the cost and time involved in running controlled user studies, many systems have emerged which use heuristics, data-driven techniques, or crowdsourcing to evaluate UIs more rapidly.
An early example is CogTool, which predicts task completion time for skilled users~\cite{cogtool}. Other tools detect underlying usability hurdles by analyzing UI layouts to find rendering errors~\cite{chen_ui_2017}, or by using crowdsourcing to find issues in interaction traces~\cite{zipt}.

Other approaches detect usability issues by modeling visual perception and highlighting mismatches with designers' expectations~\cite{guicomp}. Deep neural networks have been used to create attention maps of visual designs~\cite{predicting_visual_importance, bylinskii_learning_2017}. Our work is most similar to TapShoe, which uses a deep neural network to model users' tappability perceptions of mobile UI elements~\cite{swearngin_modeling_2019}. We extend this work by introducing a purely vision-based neural network, which enables several new applications due to its ability to run on mockups as well as functional applications. In addition, a key limitation of many automated evaluation tools is that designers must rely on their own judgment to decide how to modify their designs to improve evaluation results. Our work takes a significant step beyond prior work by using ML interpretability techniques to give designers more actionable information than predictions alone. Specifically, our system highlights the regions that influence our model's tappability predictions, and it retrieves relevant, contrasting UI examples for design inspiration.

\subsection{Interpreting and Explaining Deep Neural Network Predictions}
Deep neural networks are considered ``black box'' models since they often have too many parameters to be easily understood, and are not considered to be inherently interpretable~\cite{lipton2017mythos}.
Emergent work in the ML community has produced several algorithms and techniques that can help highlight the particular inputs to a neural network that influence its predictions.
Some methods use backpropagation to attribute pixels in an input image~\cite{ig_axiomatic}, use the convolutional features of vision models~\cite{gradcam}, or aggregate and merge highly salient pixels into regions~\cite{kapishnikov_xrai}. Other methods approximate a more interpretable, linear model to annotate what input features are near decision boundaries~\cite{lime}, or combinatorially perturb the input to determine which of its features are most influential~\cite{deepshap}. We modify the XRAI algorithm~\cite{kapishnikov_xrai} to attribute input features which influence our model's tappability predictions.

Other methods use the training dataset to provide external context that can help explain model predictions. A well-known example is to use concept vectors, which can detect the presence of learned ``concepts'' (e.g., ``stripes'', ``wheels'', or ``clouds'' in images) in a model prediction~\cite{tcav}, or identify important features across a dataset~\cite{been_ace}.
In our work, we use the latent space of our model to retrieve similar examples from our dataset, a known technique for describing model predictions by using other examples~\cite{deep_knn}. We split retrieved UIs into contrasting examples~\cite{cai_contrasting} by their tappability prediction. This exposes designers to similar UI elements with differing effects on perception, a technique based on the variation theory of learning~\cite{variation_theory}.

%% file: dataset.tex
\section{Crowdsourcing Perceived Tappability from Screenshots}

Similar to~\cite{swearngin_modeling_2019}, we perform a tappability study on a large set of UI elements in Android mobile app screens. The raters are given a screenshot from the screen set with one of the elements highlighted, and indicate whether the UI element is tappable or not. Each UI element is labeled by 5 different raters. Each worker completed up to 90 UI elements, with a median of 30.

\begin{figure}
  \centering
%   \captionsetup{width=\columnwidth}
  \includegraphics[width=\columnwidth]{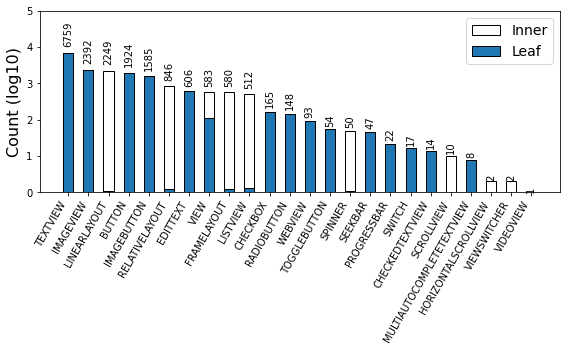}
  \caption{The type distribution for the 18,667 labeled UI elements. Blue and white splits show the proportion of leaf and inner elements in the view hierarchy.}
  \label{fig:type}
\end{figure}

We collect 18667 unique UI elements from 3218 screens from the RICO dataset~\cite{deka_rico_2017}. In the view hierarchy of each screen, we select up to five unique clickable and non-clickable elements for labeling. Similar to~\cite{zang_icons}, we asked crowdworkers to discard examples whose bounding boxes were not aligned with underlying UI elements. The same filter rules are applied as ~\cite{swearngin_modeling_2019}: we (1) choose top-level clickable elements starting from leaves and, (2) avoid choosing the children of already-chosen non-clickable elements. 

There are 24 different types of collected elements, 77\% of which are leaves in their corresponding view hierarchy trees (\autoref{fig:type}).
By analyzing the labels and the screens, we notice that some UI elements are labeled with high agreement, but others are not (\autoref{tab:agreement}).
For 44.4\% of UI elements, 5 raters agreed unanimously. However, 24.1\% of UI elements were ambiguous to raters, i.e., at most 3 agreed on a label. Nonetheless, as each element is inspected by multiple raters, our dataset has more precise labels about human tappability perception than prior work, which is desirable for machine learning tasks and data analysis. Our dataset also reveals UI elements that are indeed ambiguous, for future analysis. For model training in this work, we randomly split the dataset into 80\% of the UI elements for training, 10\% for validation to tune hyperparameters, and 10\% for testing. We release our dataset publicly on github: \dataseturl.

\begin{table}
  \caption{Agreement of tappability for the 18,667 labeled UI elements.}
  \label{tab:agreement}
  \begin{tabular}{ccc}
    \toprule
    \# of workers for agreement & \# of UI elements & ratio \\
    \midrule
    3-agreement & 4508 & 24.1\% \\ 
    4-agreement & 5872 & 31.5\% \\
    5-agreement & 8287 & 44.4\% \\
    \bottomrule
  \end{tabular}
\end{table}

% The type distribution is:
% [0, 1924, 165, 14, 606, 1585, 2392, 148, 0, 50, 17, 0, 6759, 54, 1, 0, 583, 93, 580, 2, 2249, 512, 8, 22, 846, 10, 0, 2, 47]
% The corresponding type name is: https://source.corp.google.com/piper///depot/google3/research/lux/word2act/proto/rehearsal_task.proto;rcl=387901082;l=25
% So that you can draw plot as you like.

%The label for each UI object is set to 0 for tappable, or 1 for non tappable, by a majority vote of the crowdworkers' labels.
% either an binary, which is determined by the majority of the tappable/non-tappable election, or a float value in \{.0, .2, .4, .6, .8, 1.\}, which the the ratio of raters who choose tappable. %@eschoop you may add how you use it here. 
%We randomly split the dataset into 80\% of the UI elements for training, 10\% for validation to tune hyper parameters of the models and 10\% for testing. 

% original label file: 
% /cns/is-d/home/word2act/rl/rico3k_tappability/tap_relabel_merge_screen3462_obj20160.csv

% worker number 2547 raters, each workers contributes 30 labels (mid).

%% file: model.tex
\begin{table*}
  \caption{Tappability prediction model performance on our new dataset. Our model, which only uses pixels, clearly outperforms~\cite{swearngin_modeling_2019} when only run on pixels. It has slightly higher AUC, similar precision, and slightly lower recall compared to the all-features replicated model.}
  \label{tab:benchmarks}
  \begin{tabular}{lcccc}
    \toprule
    Model & AUC & Precision (\%) & Recall (\%) \\
    \midrule
    % Ours (pixels only) & 0.9026 & 90.71 & 75.85 \\
    Ours (pixels only) & 0.9030 & 91.54 & 80.23 \\
    Swearngin et al.~\cite{swearngin_modeling_2019}; pixels + all other features & 0.8437 & 91.65 & 84.53 \\
    Swearngin et al.~\cite{swearngin_modeling_2019}; pixels only & 0.6521 & 76.79 & 80.79 \\
    \bottomrule
  \end{tabular}
\end{table*}

\section{Modeling Perceived Tappability from Images}

Since the applications in our dataset use many UI frameworks and design styles, the patterns persistent in this data can be generalized to predict the tappability of elements in many kinds of mobile UIs. In this section, we describe how we use our dataset to train a Convolutional Neural Network (CNN) model for tappability prediction.
The problem statement for our model is: given an input screenshot and region of interest (a rectangular area within the input screenshot), predict whether or not users will perceive the indicated UI element as tappable or not tappable.

Our CNN model is purely vision-based, which significantly differs from prior work in tappability prediction~\cite{swearngin_modeling_2019}, and provides several advantages.
While earlier tappability prediction models required multiple feature types as input (e.g., a screenshot and a selected element's Android View type, text content, and its intended tappability), our model only uses screenshot pixels as input. This significantly broadens the set of applications our model may be used for, such as UIs that are not fully-specified. For example, designers may be able to use our model to evaluate iterations in earlier design stages since it can operate on visually realistic mockups.
However, since our model does not directly capture text, element type, or intended clickability information from input UIs, the model from Swearngin et al.~\cite{swearngin_modeling_2019} may have advantages in contexts where non-visual signifiers (e.g., text content) are used by designers to explicitly indicate tappability.
Since our vision-based model does not rely on platform-specific inputs (i.e., element types), it can be fine-tuned for platform-agnostic applications. This also makes it easier to adapt our model to other domains in future work, such as emergent datasets of iOS applications~\cite{wu_screen_parsing}, or other downstream tasks, e.g., predicting accessibility barriers~\cite{screen_recognition_apple}.

% \begin{figure}
%     \centering
    % \includegraphics[trim={0 0 20px 0},clip,width=0.8\textwidth]{figures/tap_model_embeds.pdf}
%     \caption{Our model accept inputs of a 960 by 540, 3-channel image plus a corresponding binary mask channel, where a rectangular patch of $1$s in the mask specifies the UI region of interest. The binary mask is concatenated to the input image along the channel dimension as a fourth channel, and is fed directly into our classifier.}
%     \label{fig:model}
%     \Description{TODO}
% \end{figure}

Our model's inputs are specified as follows. Let $I \in \mathbb{R}^{h\times w\times 3}$ denote the pixel values of a UI screenshot, where
$h$ and $w$ are the screen height and width, and 3 is the number 
of channels (i.e., RGB).
Let $(x_{min}, y_{min})$ and $(x_{max}, y_{max})$ denote the top-left and bottom-right corner 
coordinates of a target UI element bounding box respectively.

A naive implementation of using CNNs for learning tappability is to crop the target element's pixels from $I$ and feed them to a CNN.
However, this discards important contextual information in the screen, making it difficult to learn an effective model. Instead, we feed the entire RGB screenshot to the model along an additional mask channel in the input.
For a given element, we first create a binary mask $M \in \{0, 1\}^{h\times w}$, using $i$ and $j$ as row and column indices, respectively:
$$
M_{ij} = 
    \begin{cases}
    1, & \text{if}\ y_{min} <= i < y_{max}\ \text{and}\ x_{min} <= j < x_{max}\\
    0, & \text{otherwise}
    \end{cases}
$$
In other words, the entries corresponding to the target element's pixels are 1's and 
the others are 0's in the binary mask.
We then concatenate $I$ and $M$ along the channel dimension to form the input to the 
model: $I'=[I, M]$ of shape [$h, w, 4$].
To the model, $I$ provides pixel information of the whole screen, while $M$ indicates 
the screen area for which the model should predict tappability (\autoref{fig:teaser}).

Specifically, our model is a Resnet-18~\cite{resnet_he}, modified to accept a larger input image with a dimension of 960 by 540 (to accommodate mobile UI screenshots) along with the corresponding binary mask. The model outputs softmax probabilities for two classes: tappable, or not tappable. We train our model on the training set by minimizing cross-entropy loss, using Stochastic Gradient Descent with Nesterov momentum, with a learning rate of 0.05 and a batch size of 1024, for 1500 epochs. Our learning rate decayed by an order of magnitude (dividing by 10), after epochs 100, 500, 1000, and 1300. 
We evaluated how well our model predicted user perceptions of the tappability of UI elements with our test set.
Our model achieved a precision of 91.54\% and recall of 80.23\% with a decision threshold of 50\%, and AUC of 0.9030.

To compare the performance of our model to previous work in tappability prediction, we replicated the model from Swearngin et al.~\cite{swearngin_modeling_2019} and benchmarked this model on our new dataset in two separate configurations: by using all of its input features (screenshot pixels, region pixels, component text, component type, and intended tappability), and by using pixels only (from the screenshot and region).
Our model, which only uses pixels, clearly outperforms the replicated model~\cite{swearngin_modeling_2019} when it only runs on pixels. When the replicated model uses all input features, including those from the view hierarchy, on our dataset, our model achieves better AUC and similar precision, but has slightly lower recall when using a 0.5 decision boundary (\autoref{tab:benchmarks}).
% We found the AUC of our model is higher than the replicated all-features model, which shows that visual features alone are extremely powerful in predicting users' perceptions of tappability.
% While our model's precision is comparable to the all-features model, its recall is just slightly below the pixels-only replicated model.
The slightly lower recall of our model is likely due to the distribution of tappable elements in our dataset, which can be addressed by fine-tuning the decision threshold.

%% file: explanations.tex
\section{Explaining Tappability Predictions}

\begin{figure}
    \centering
    \includegraphics[width=0.9\linewidth]{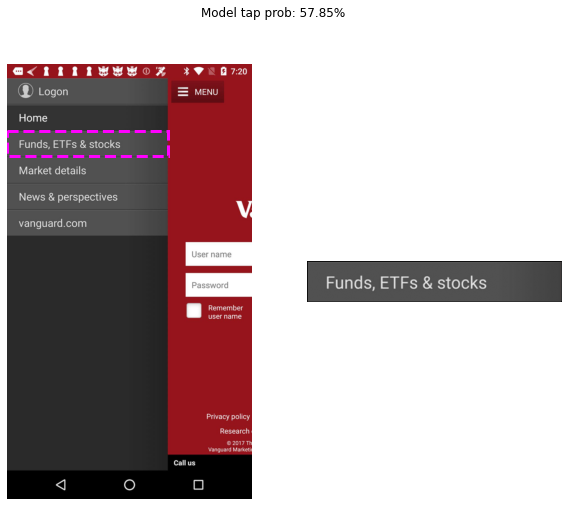}
    \caption{The input to our model as a running example to this section, a randomly selected screenshot from our dataset. The element of interest is indicated as a magenta dashed rectangle. Our model predicts the element is tappable with a probability of 57.85\%.}
    \label{fig:example_input}
    \Description{TODO}
\end{figure}

Our neural network can be used to model users’ perceptions of tappability for a broad variety of mobile UI elements.
However, the predictions of models like ours are limited in the sense that designers must rely on their own judgment to determine what visual cues were responsible for the prediction, and, if needed, how the design must be modified to improve its perception (\autoref{fig:example_input}).
We draw upon techniques from XAI and ML interpretability to provide deeper explanations of our model's predictions, both in the context of the input itself, as well as examples the model has learned from. We implement two types of explanations: at the \textit{local} level, to suggest which elements in the input screenshot were most influential, or ``salient'', to a given prediction, and at the \textit{global} level, to show how other applications with similar design patterns can influence tappability perception positively and negatively.

\begin{figure}
    \centering
    \includegraphics[width=\linewidth]{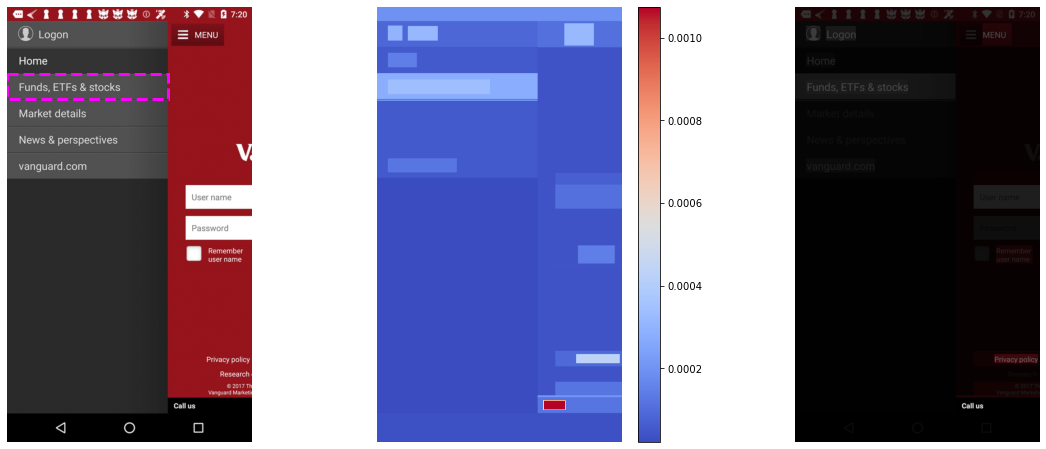}
    \caption{\textbf{Left:} The same input screenshot as in \autoref{fig:example_input} with a selected element in a magenta dashed rectangle. \textbf{Center:} the heatmap generated by XRAI, using regions from UI elements. The regions which most strongly influence the selected element's tappability prediction are rendered in red, while the least influential regions are rendered in blue. Some text is extremely highly attributed (an anomaly). \textbf{Right:} the input screenshot filtered by the values of the saliency heatmap. The elements most important to the tappability prediction are the brightest.}
    \label{fig:example_xrai_3}
    \Description{TODO}
\end{figure}

\begin{figure}
    \centering
    \includegraphics[width=\linewidth]{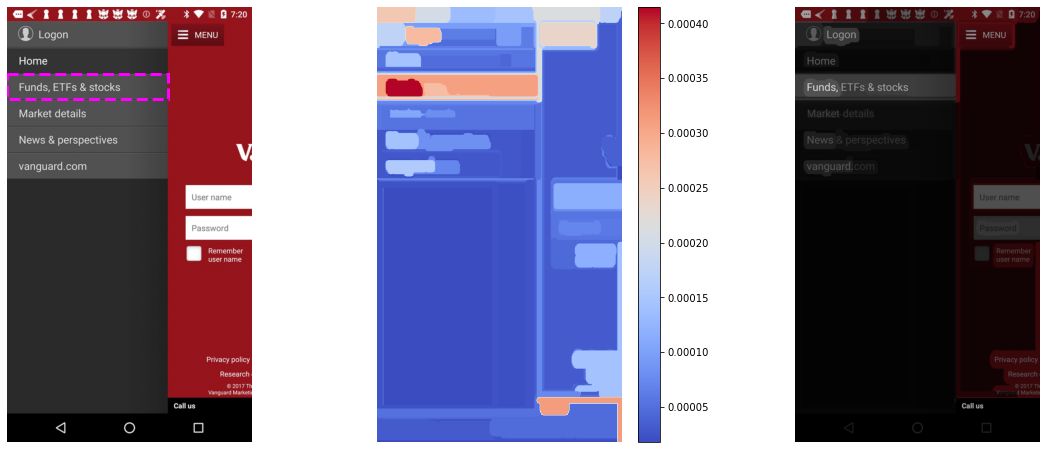}
    \caption{\textbf{Center:} the same XRAI calculation as in \autoref{fig:example_xrai_3}, but without using provided regions from the UI element bounding boxes. Regions are generated using Felzenszwalb segmentation.}
    \label{fig:example_xrai_fz}
    \Description{TODO}
\end{figure}

\subsection{Attributing Tappable UI Elements with Saliency Techniques}

To provide a \textit{local} explanation of the model's predictions, we use the XRAI algorithm~\cite{kapishnikov_xrai}, a gradient-based algorithm which produces a heatmap highlighting what regions of an input image were the most influential to a given model output, also known as a saliency map (\autoref{fig:example_xrai_3}).
Importantly, while the output of saliency algorithms like XRAI are correlative, and cannot explain the causal reasons behind model predictions, they are often useful for gaining a better understanding of model behavior in many applications~\cite{been_moritz_saliency_checks}.
In our use case, we use XRAI to generate a heatmap of the UI components in a mobile app screenshot that most strongly influence the tappability prediction for a particular element.
XRAI calculations and heatmaps are particular to the specified UI element in a tappability prediction, since predictions for different UI elements can depend on their particular context and relationship to other UI elements.
Designers can use the XRAI heatmap to see when the perceived tappability of a particular element is heavily influenced by other regions on the screen, e.g., how introducing a new component changes the perception of surrounding elements.

The XRAI algorithm works by first oversampling the input image into overlapping superpixels of different sizes. Next, Integrated Gradients, a pixel-based attribution method~\cite{ig_axiomatic}, is calculated on the input image from black and white baselines. These pixel-level attributions are then aggregated by summing over segments, ranking segments from most to least important, and merging them up to a selected threshold.
We make one key modification to the XRAI technique. Rather than oversample the image using Felzenszwalb segmentation, we use the native bounding boxes of mobile UI elements if they are available or can be specified. This means we can directly summarize model attributions for regions corresponding to mobile UI widgets, reducing the noise from automated segmentation methods (\autoref{fig:example_xrai_fz}).

% \deleted{For our implementation, we provide the screenshot as an input, and set the binary mask to all 1's, selecting the entire screenshot as the ``region of interest'' for purposes of running XRAI. We found that selecting a smaller bounding box tends to clip gradients, resulting in a much less useful heatmap.}

\subsection{Explaining Predictions with Similar Contrasting Examples}

\begin{figure}
    \centering
        \centering
        \includegraphics[width=\columnwidth]{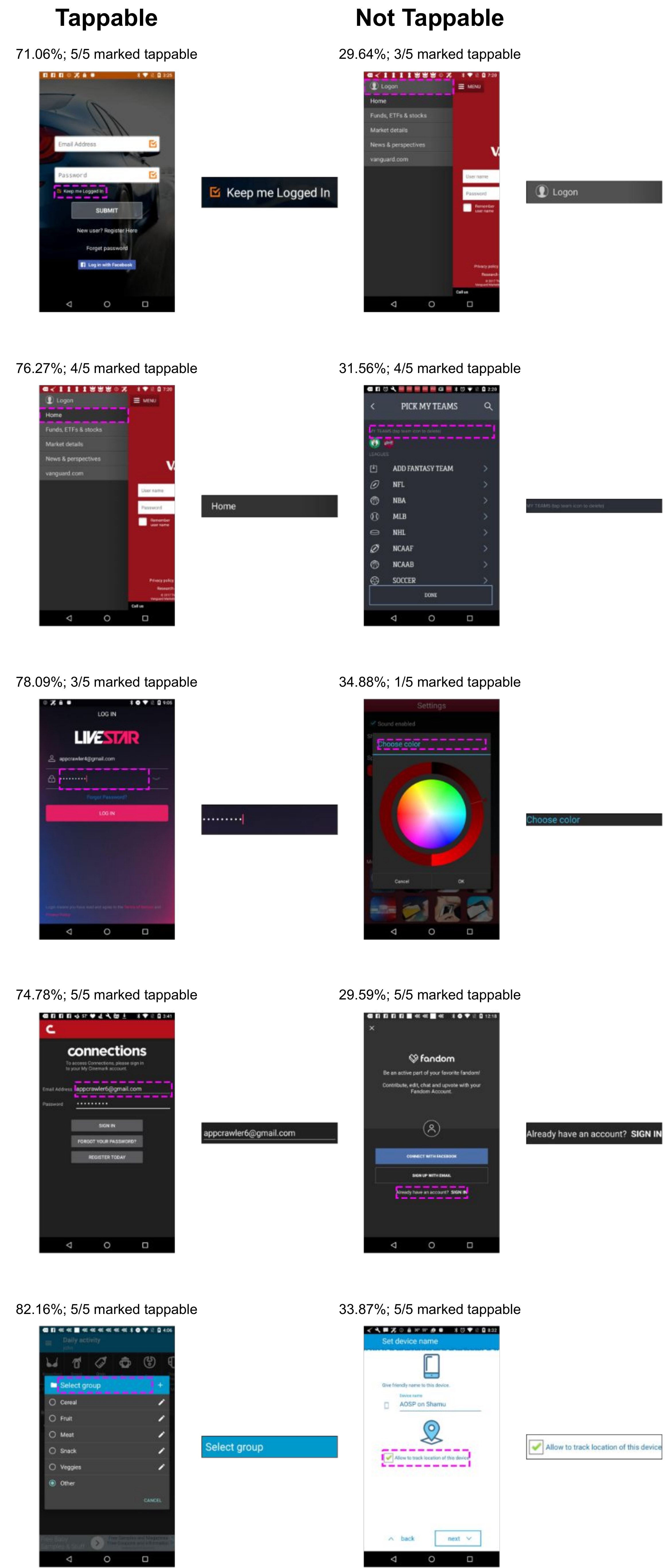}
        \caption{
        % Two examples of selection methods for nearest neighbors.
        % \textbf{Left set of two columns:}
        Nearest neighbors to the input screenshot from \autoref{fig:example_input}, split by tappability predictions. Examples the model predicts as tappable are on the \textbf{left}, with non-tappable examples on the \textbf{right}. Examples contain an entire screenshot with a specified region, and the region in a larger view. Columns are sorted by distance to the original input in the model's latent space (most similar on top).
        % \textbf{Right set of two columns:} Nearest neighbors split by a tappability vote. Crowdworkers' labels are each one vote, and the model's non-thresholded prediction is worth four votes. This allows unanimous consensus by crowdworkers to override the model's prediction. The voting method can sometimes reduce the noisiness of suggestions. Note the top left example is the same as the input image, as expected, since the input was sourced from the same dataset and has a distance to itself of 0.
        }
        \label{fig:example_nns}
        \Description{TODO}
\end{figure}

Our \textit{global} explanation method situates the given prediction in the context of retrieved examples from our mobile UI dataset. We use nearest neighbors on embeddings from our model to find examples the model considers similar. The model's embeddings capture visual similarity, and the rough position and size of the input bounding box (see~\autoref{sec:analysis}). These nearest neighbors are then split by the model's tappability prediction, creating a contrasting explanation~\cite{cai_contrasting}---a visualization of a set of UIs that have similar designs to the input, but opposing influences on the perception of tappability. This acts as a set of curated examples for design inspiration to help designers make changes that affect users' tappability perceptions of UI elements (\autoref{fig:example_nns}).

To capture embeddings from our neural network, we take the output from its final convolutional layer and flatten it into a 512-dimensional vector. We precompute embeddings for every mobile UI example in our source dataset, and split them into two separate indexed arrays of predicted tappable and nontappable examples. To filter out potentially confusing or ambiguous examples, we limit these lists to examples which have >65\% and <35\% tappability probabilities, per our model's predictions. In practice, we found that splitting based on model predictions produces more consistent results than ground-truth human labels. We use the \texttt{NearestNeighbors} learner from the \texttt{sklearn} Python package to search for the 5 nearest neighbors from each list (showing 10 examples total), to embeddings from an input image.

%% file: analysis.tex
\section{Analysis of Selected Examples}
\label{sec:analysis}

In this section, we sample real-world screenshots from our dataset to show how our model performs and what our explanations capture. We randomly select four elements from our dataset that have associated regions corresponding to common Android UI leaf elements: \texttt{ImageView}, \texttt{Button}, \texttt{TextView}, and \texttt{EditText}. For each of these inputs, we show the output of our model and explanation methods, and describe what could be inferred about the behavior of our model. In \autoref{sec:discussion}, we summarize trends apparent in our model across examples and discuss their implications and opportunities for future work.

\begin{figure}
    \centering
    \includegraphics[width=\columnwidth]{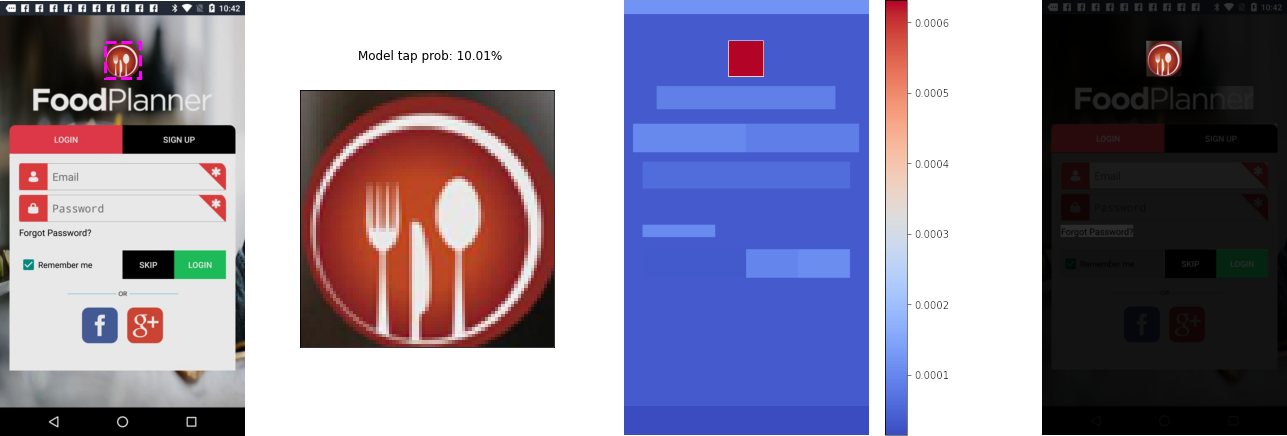}
    \caption{Example from \autoref{sec:example_imageview}. \textbf{Left:} input screenshot with a \texttt{ImageView} element selected, annotated with a magenta dashed rectangle.
    \textbf{Center Left:} Close-up view of the selected UI element. The model predicts it is not tappable, with a 10.01\% tappability probability.
    \textbf{Center Right:} The XRAI heatmap most strongly illuminates the selected element, and does factor in other elements significantly.
    \textbf{Right:} the input screenshot filtered by the values of the saliency heatmap.}
    \label{fig:food_imageview_analysis}
    \Description{TODO}
\end{figure}

\begin{figure}
    \centering
    \includegraphics[width=\columnwidth]{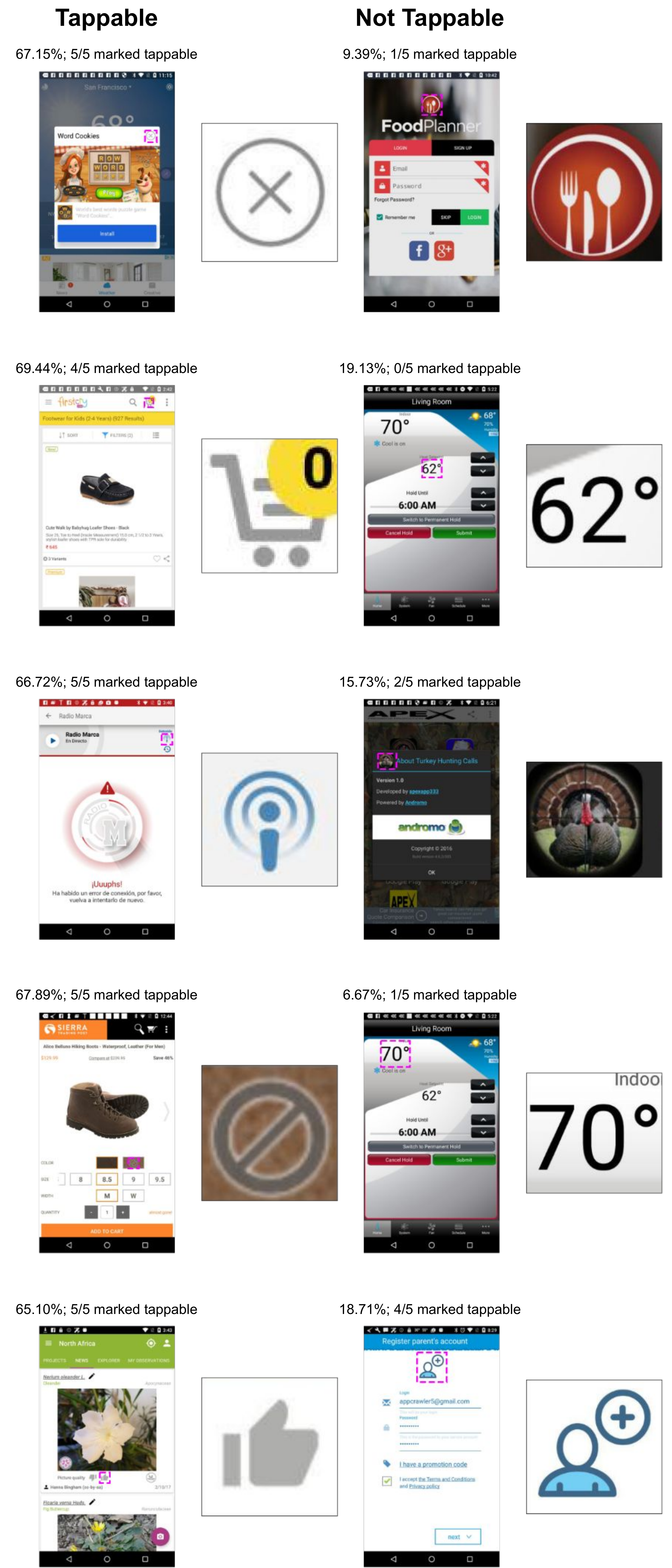}
    \caption{Nearest neighbors from \autoref{sec:example_imageview}, split by thresholded model predictions (tappable neighbors on the left). Many neighbors (both tappable and not) are graphical (icons and drawings). Tappable elements tend to be smaller, and situated near the edges of other elements. Non tappable elements are generally larger, and closer to the center of the screen.}
    \label{fig:food_imageview_nns}
    \Description{TODO}
\end{figure}

\subsection{\texttt{ImageView}: Food App Header Logo}
\label{sec:example_imageview}

This randomly-selected UI and element is a screenshot from a food application, presenting a complex login view with many clickable buttons and graphics. The selected region paired with this UI screenshot is a logo placed above the login form. The model predicts this element is not tappable, with a 10.01\% tap probability.

The XRAI heatmap strongly attributes the input element as important to its tappability prediction, and does not factor other elements in the screen much. It is possible that, since the input element is a graphic, the model does not consider surrounding elements a significant factor. Combined with the relatively high model confidence, we can assume that properties of the input element itself (its appearance or position) strongly signify non-tappability on their own. This means that making significant changes to other elements on the screen would likely not impact the perceived tappability of this element.
% The XRAI heatmap strongly highlights a ``Forgot Password?'' link and the selected logo (\autoref{fig:food_imageview_analysis}). Other elements (e.g., \texttt{EditText} views and buttons) are highlighted as well, but with less intensity. Similar to the XRAI results from the finance app, the model prediction for the password reset element does not match its high attribution---the model predicts it is not tappable, with a 16.84\% tap probability. In this example, the high attribution values may signify that the logo and text influence the tappability of \textit{nearby} elements, e.g., as an icon next to text, and a descriptor of a text field.

Most nearest neighbors of the food app also contain graphical elements and icons, with the exception of large text objects that have similar locations and sizes on the screen as the input (\autoref{fig:food_imageview_nns}). The non tappable elements are generally larger, and closer to the center of the screen, matching the style of the input. Tappable elements tend to be icons commonly associated with actions, e.g., a shopping cart and an ``X'' to close a dialog.

\subsection{\texttt{Button}: Health App Card Button}
\label{sec:example_button}

\begin{figure}
    \centering
    \includegraphics[width=\columnwidth]{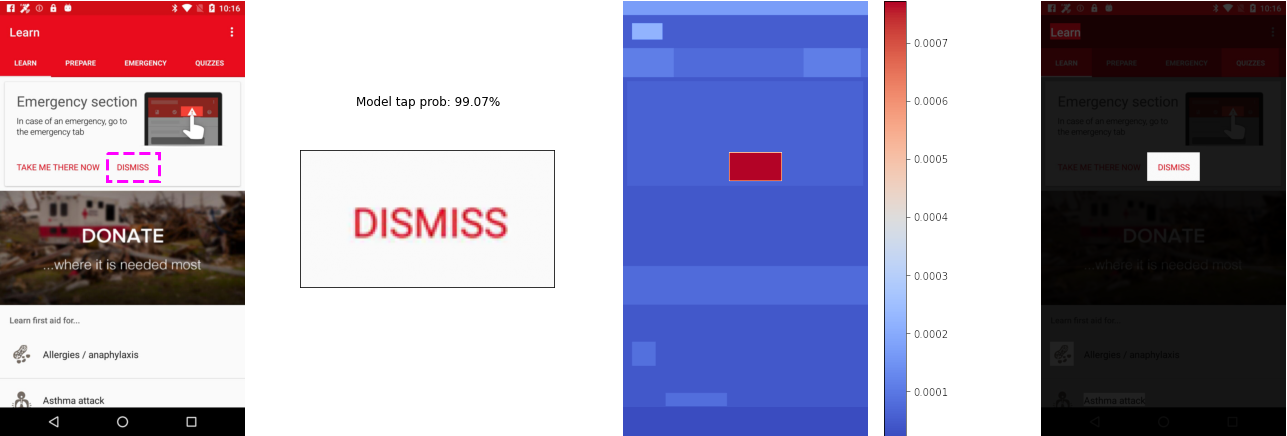}
    \caption{Example from \autoref{sec:example_button}. \textbf{Left:} input screenshot with a \texttt{Button} element selected, annotated with a magenta dashed rectangle.
    \textbf{Center Left:} Close-up view of the selected UI element. The model predicts it is tappable, with a 99.07\% tappability probability.
    \textbf{Center Right:} The XRAI heatmap most strongly illuminates the input element itself. This is likely because the model has learned to associate Material Design buttons with perceptions of tappability, and does not need to reference much context to establish a confident prediction.
    % \textbf{Center Right:} The XRAI heatmap most strongly illuminates the header text of a card, the header text of the app, and the first text element, adjacent to an icon, in a \texttt{ListView}. Buttons and other elements in the lower \texttt{ListView} are highlighted as well, but with less intensity.
    \textbf{Right:} the input screenshot filtered by the values of the saliency heatmap.}
    \label{fig:emergency_button_analysis}
    \Description{TODO}
\end{figure}

\begin{figure}
    \centering
    \includegraphics[width=\columnwidth]{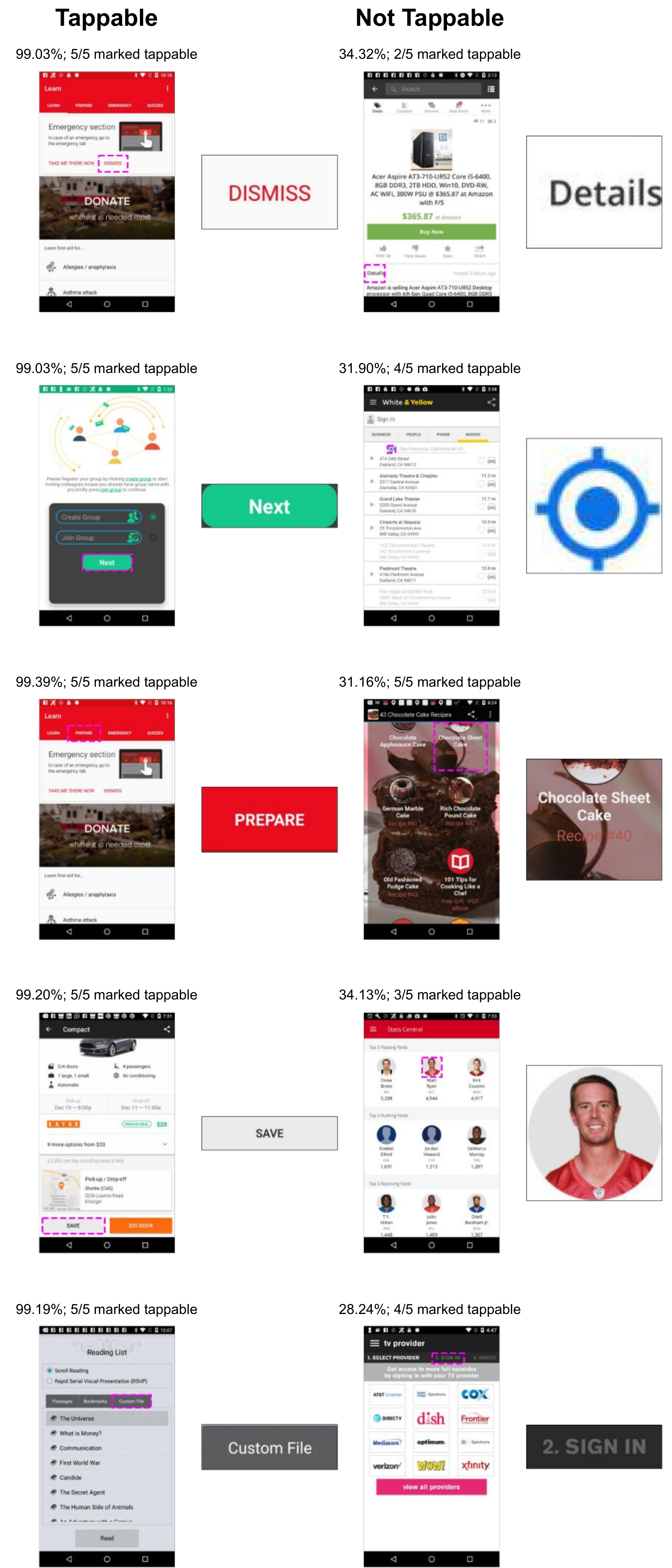}
    \caption{Nearest neighbors for \autoref{sec:example_button}, split by thresholded model predictions (tappable neighbors on the left). While many tappable elements contain similarly-styled buttons from other apps, the non-tappable elements display highly varied elements, with and without text. A potential cause of this is that most elements near the input button in latent space are other buttons; and the nearest non-tappable examples are significantly further away, so they are not as visually similar.}
    \label{fig:emergency_button_nns}
    \Description{TODO}
\end{figure}

This element is a screenshot from a health application, presenting a complex view with a card, image, and list. The selected region paired with this UI screenshot is a ``Dismiss'' button within a card. The model predicts this element is tappable, with a 99.07\% tap probability (\autoref{fig:emergency_button_analysis}).

Similar to the \texttt{ImageView} example, the XRAI heatmap most strongly illuminates the input element itself. This is likely because the model has learned to associate Material Design buttons with a strong perception of tappability, and does not need to reference much context to establish a confident prediction. The attributed text in the screen's title card (``Learn'') may suggest the model's attention to a common Material UI standard.
% Similar to the other examples, the XRAI heatmap strongly attributes text elements: the app header, the title of the card (``Emergency section''), and the first element of the list at the bottom. Buttons and tabs are also highly attributed, and each contain text elements. This further suggests that small text snippets, particularly bolded or colored text, could signify the tappability of their parent elements.

Tappable neighbors are entirely buttons and tabs with overlaid text and high tappability scores. Non-tappable neighbors are more mixed, including descriptive text, icons, and even images. The predictions for several non-tappable examples disagree with the underlying raters' labels (\autoref{fig:emergency_button_nns}). The noise in the non-tappable examples could be a limitation of discretizing the neighbors by tappability prediction, an effect discussed further in \autoref{sec:discussion_nns}.

\begin{figure}
    \centering
    \includegraphics[width=\columnwidth]{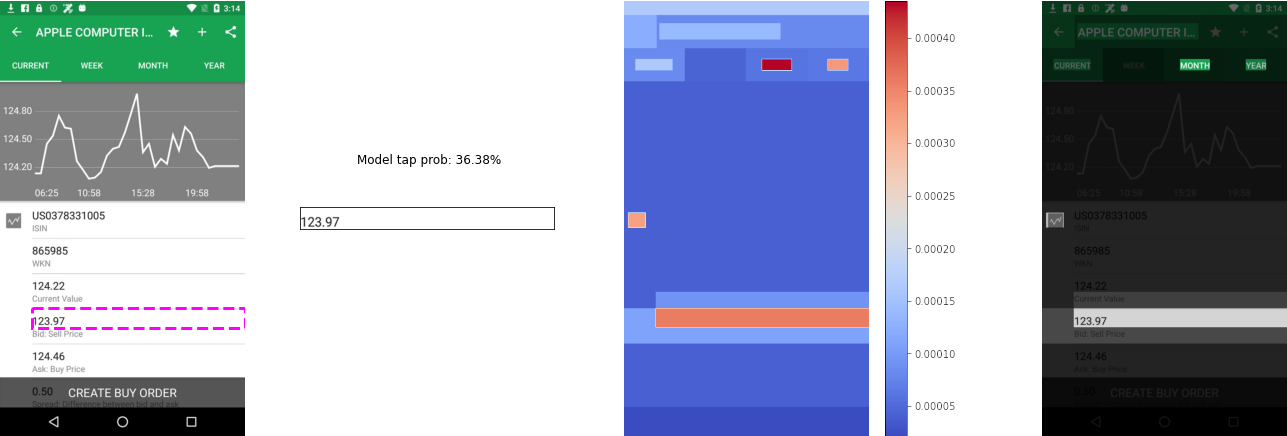}
    \caption{Example from \autoref{sec:example_textview}. \textbf{Left:} input screenshot with a \texttt{TextView} element selected, annotated with a magenta dashed rectangle.
    \textbf{Center Left:} Close-up view of the selected UI element. The model predicts it is not tappable, with a 36.38\% tappability probability.
    \textbf{Center Right:} The XRAI heatmap strongly illuminates the text view region itself, while also strongly highlighting text in tab navigation and an icon adjacent to a nearby text view.
    \textbf{Right:} the input screenshot filtered by the values of the saliency heatmap.}
    \label{fig:stocks_textview_analysis}
    \Description{TODO}
\end{figure}

\begin{figure}
    \centering
    \includegraphics[width=\columnwidth]{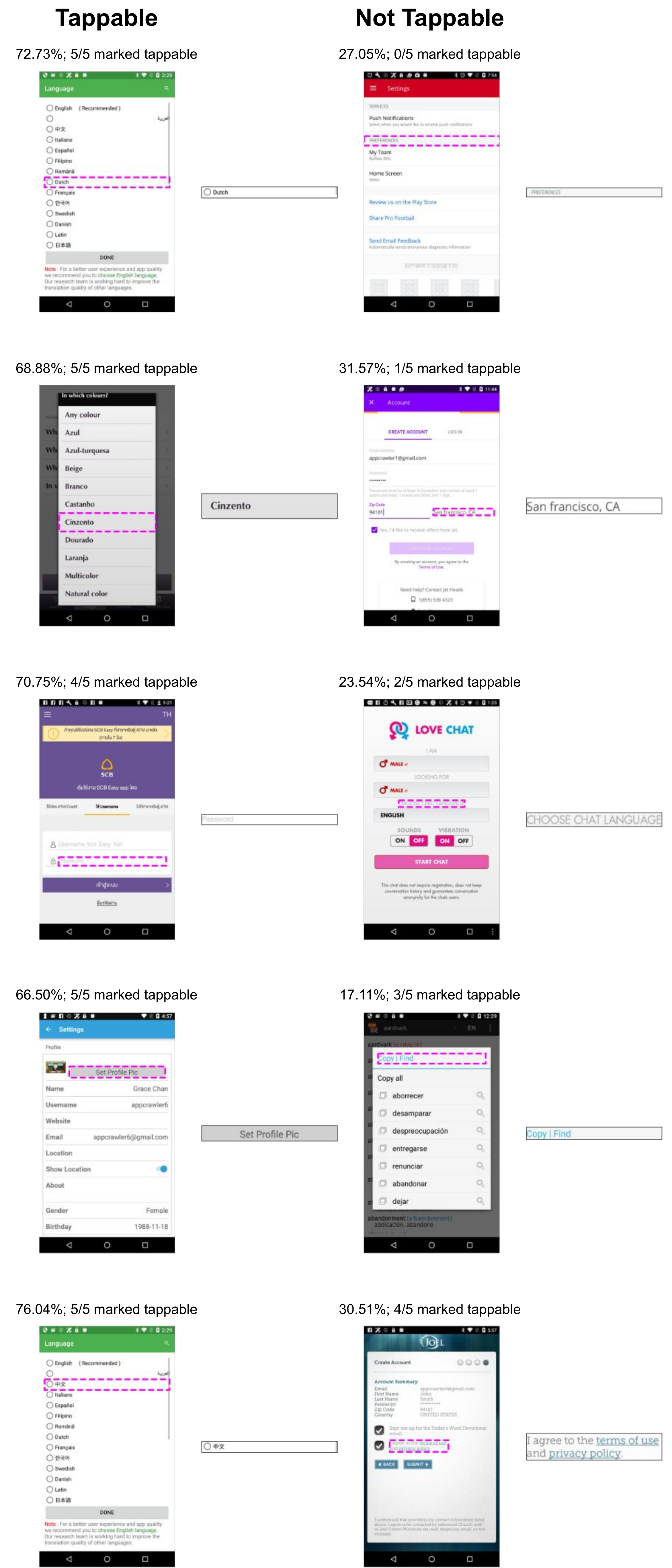}
    \caption{Nearest neighbors from \autoref{sec:example_textview}, split by thresholded model predictions (tappable neighbors on the left). Tappable elements are similar to the input, in that most contain stacked text elements (two with a green header bar).}
    \label{fig:stocks_textview_nns}
    \Description{TODO}
\end{figure}

\subsection{\texttt{TextView}: Finance App List Item}
\label{sec:example_textview}
This example is a screenshot from a finance app, with a view presenting a chart and pricing details of a stock (\autoref{fig:stocks_textview_analysis}). The selected region paired with this UI screenshot is a text field displaying a bid price. The model predicts this element is not tappable, with a 36.38\% tap probability.

The XRAI heatmap strongly illuminates the region itself, while also strongly highlighting text in tab navigation and an icon adjacent to a nearby text view.
Although the input region is generally expected to be the most important element for its own prediction, one element of tab text is highly attributed, an anomaly. This may be due to variances in \texttt{TextView} tappability when below navigation tabs.
% , or possibly an artifact of how XRAI normalizes region attributions by their size~\cite{kapishnikov_xrai}.
It is also worth noting that surrounding text views are lightly attributed as well, suggesting the model has factored some surrounding context into the input element's prediction.
% Of note, the icons in the top right of the XRAI heatmap do not appear to be highly attributed. However, this does not necessarily mean they would be predicted as non tappable (e.g., the ``share'' icon is predicted as tappable, with a probability of 97.97\%). This is because the saliency attribution task is not equivalent to tappability prediction. This effect and its implications are discussed further in \autoref{sec:discussion}.

All nearest neighbors of this input screenshot share strong visual similarities with the input (text on a light background), but appear in different contexts (\autoref{fig:stocks_textview_nns}).
% , as filtered by thresholded\footnote{Predictions with a tappability probability of under 35\% are considered non tappable, and over 65\% tappable, to reduce potentially confusing examples.}model predictions .
Many tappable elements have icons or graphics nearby, which possibly serve as signifiers of the tappabillity of the adjacent text. Non tappable elements have brighter text, and are often placed as descriptions next to tappable elements.
It is worth noting that many tappable elements are also \texttt{ListView}s, 2 of which have similar color schemes, indicating the model is factoring multiple contextual elements within the input example besides the region itself.

\subsection{\texttt{EditText}: Entertainment App Login Field}
\label{sec:example_edittext}

\begin{figure}
    \centering
    \includegraphics[width=\columnwidth]{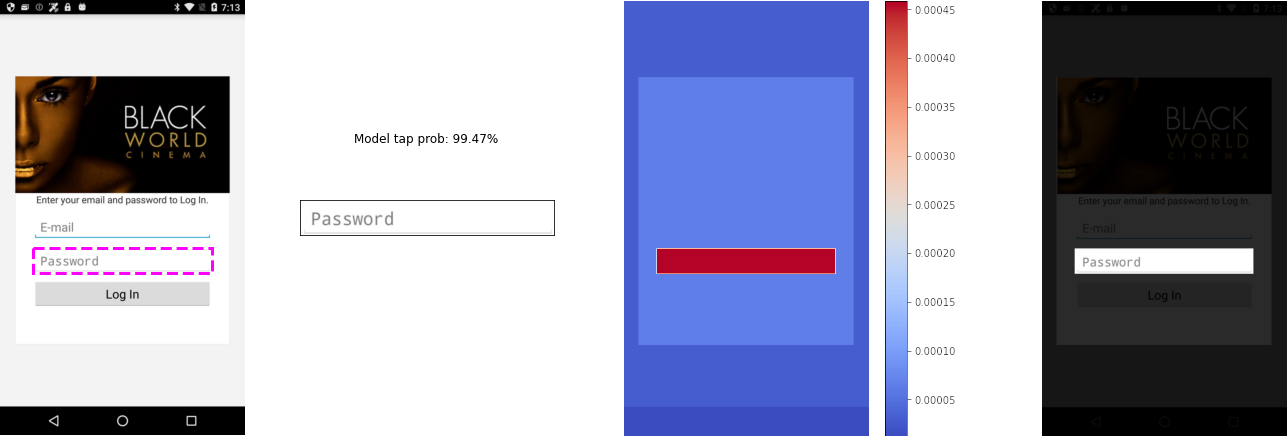}
    \caption{Example from \autoref{sec:example_edittext}. \textbf{Left:} input screenshot with a \texttt{EditText} element selected, annotated with a magenta dashed rectangle.
    \textbf{Center Left:} Close-up view of the selected UI element. The model predicts it is tappable, with a 99.47\% tappability probability.
    \textbf{Center Right: } The XRAI heatmap strongly attributes the \texttt{EditText} view, and does not attribute other elements on the screen.
    % \textbf{Center Right:} The XRAI heatmap most strongly illuminates a small text description of the login screen. The ``Log In'' button and ``Password'' text field are also highlighted.
    \textbf{Right:} the input screenshot filtered by the values of the saliency heatmap.}
    \label{fig:login_edittext_analysis}
    \Description{TODO}
\end{figure}

\begin{figure}
    \centering
    \includegraphics[width=\columnwidth]{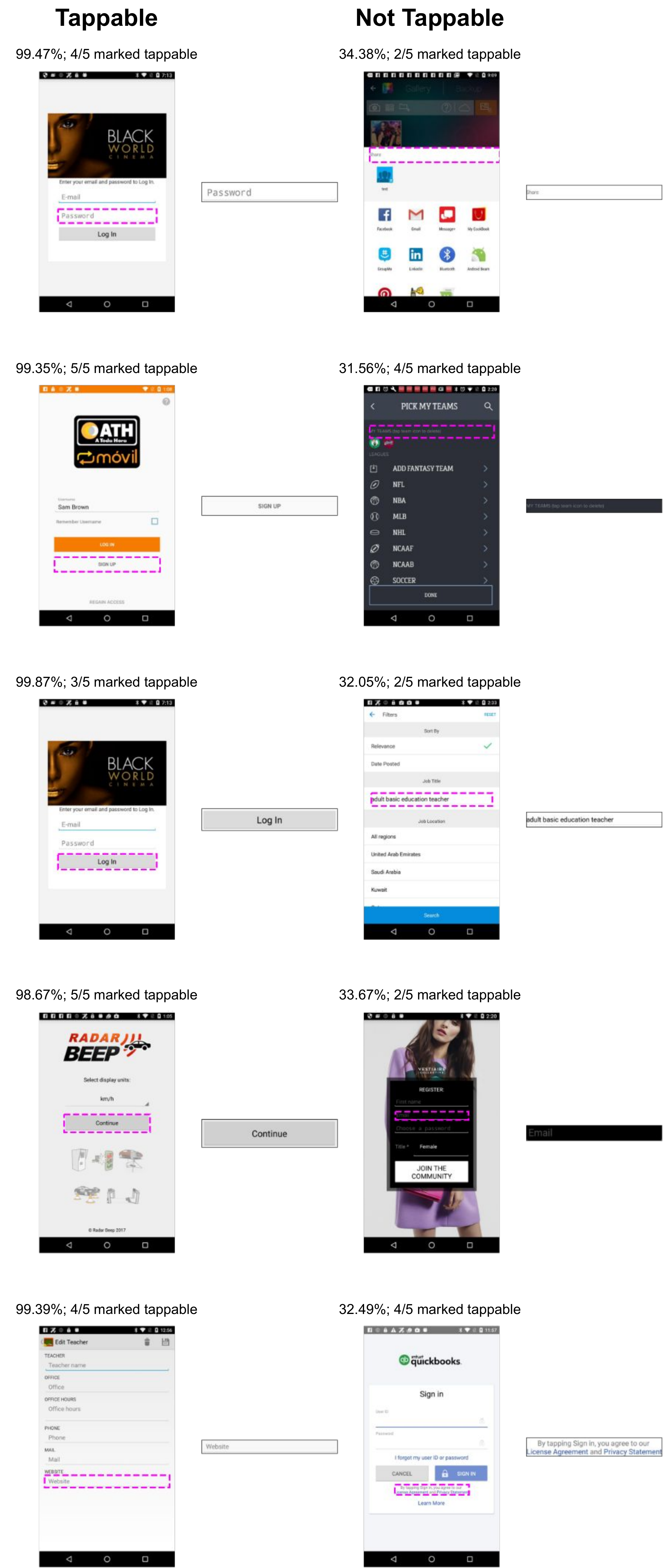}
    \caption{Nearest neighbors from \autoref{sec:example_edittext}, split by thresholded model predictions (tappable neighbors on the left). The tappable neighbors are highly visually similar, although they are not all the same input type as the input (many are buttons with a light background).}
    \label{fig:login_edittext_nns}
    \Description{TODO}
\end{figure}

This element is a screenshot from an entertainment application, a simple login view. The selected region paired with this UI screenshot is a ``Password'' text field. The model predicts this element is tappable, with a 99.47\% tap probability (\autoref{fig:emergency_button_analysis}).

Like previous examples, the XRAI heatmap strongly attributes the selected \texttt{EditText} view, and does not attribute other elements on the screen. Similar to the \texttt{Button} example, it is possible that the model has learned an association between the Material UI \texttt{EditText} component and strong perceptions of tappability.

% This XRAI heatmap is a good case study of how small text is often highly attributed with our saliency method, although the text fields and ``Log In'' button are also attributed (\autoref{fig:login_edittext_analysis}).
The image in this view does not appear to be attributed differently from the entire login card. While it is likely that the model determined the image is not a signifier of tappability, it could also be, in part, due to these two being the same actual element in the source UI view hierarchy. Our use of XRAI is limited by the bounding boxes provided from the source UI view structure---large objects in UIs may cause XRAI to aggregate too much detail from pixel attributions beneath. In practice, this may not be a significant limitation, since many large UI objects inherit a single tappability attribute.

% Nearest neighbors of the entertainment app UI are also split by thresholded tappability predictions.
Like the \texttt{TextView} example, the tappable neighbors are all visually similar, with text over a light-colored background, comprising buttons and text fields. Non tappable neighbors are, similarly, text elements in different contexts: descriptions of nearby objects, instructions, or hyperlinks. Of note, the third non-tappable neighbor is also a \texttt{EditText} element. A probable distinguishing feature of this element is that the text is dark (not grayed), and thus the model could be confusing this element for a text description (\autoref{fig:login_edittext_nns}).

%% file: eval.tex
\section{Exploratory Evaluation with Professional Designers}
\label{sec:eval}

To better understand how our model and its explanation outputs can be used in design practice, we conducted an exploratory evaluation with professional UI/UX designers, and analyzed the successes and drawbacks of our approach.

\begin{figure}
    \centering
    \includegraphics[width=\columnwidth]{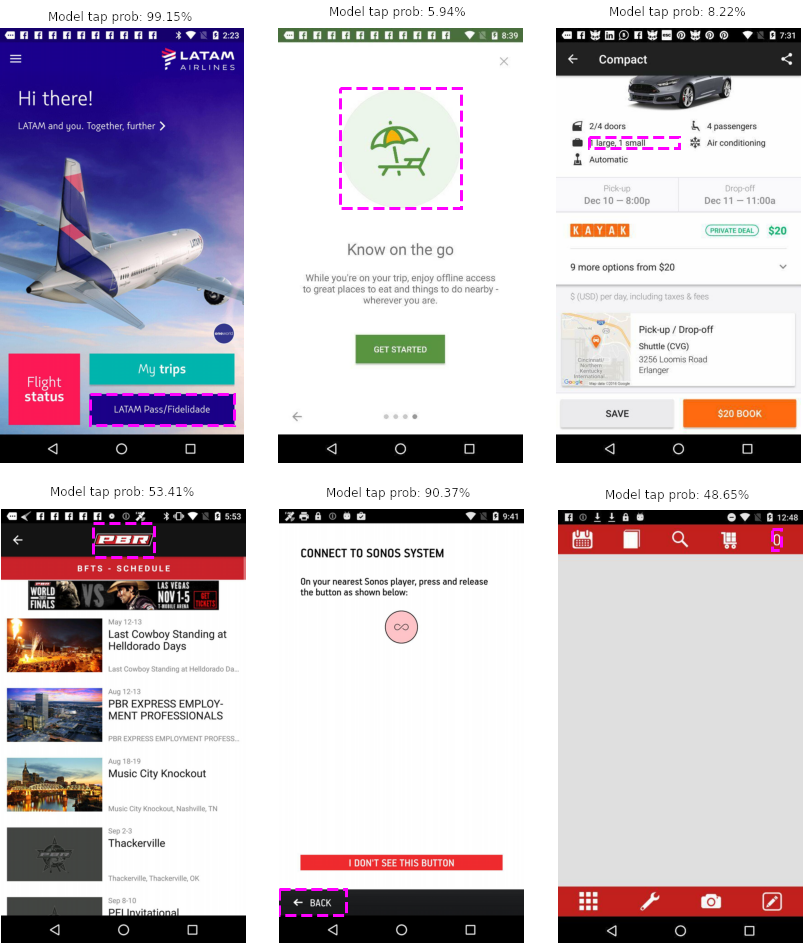}
    \caption{The six mobile UI screenshots with preselcted UI elements used in our user evaluation. Selected elements are circled with a dotted magenta line. The top left UI screenshot was used for onboarding participants.}
    \label{fig:user_study_UIs}
\end{figure}

\subsection{Participants \& Study Design}
We recruited \added{14} participants at a large technology company. We excluded one participant's results from analysis because they did not submit any written feedback. These participants were from multiple teams and had an average 11 years (standard dev. 7.5 years) of professional UI/UX design experience. To capture a variety of scenarios, we randomly selected six UI examples (mobile app screenshots with a preselected UI element) from our dataset for review. Input UIs and predictions are shown in \autoref{fig:user_study_UIs}, outputs from explanation algorithms are shown in \autoref{sec:user_study_figs}. Examples counterbalanced prediction (tappable/non tappable), prediction confidence (high: $> 0.85$; low: $< 0.15$), and  rater (worker) agreement (high: 5-rater agreement; low: split 2-3 in either direction). We selected the tappable/high confidence/high agreement example for use in onboarding. Tappability labels from raters associated with the examples were not shown to participants. Our study plan was reviewed by our company's legal and privacy boards, and participants were required to give informed consent before trials.

During each session, we first described our model and its explanation outputs using the onboarding example. Then, the five remaining examples were shown in a randomized order, together with the model's outputs. For each example, we asked our participants to think out loud; reflect on whether they understood or agreed with the model's outputs; and suggest how the selected element in the example could be altered to influence its perceived tappability.
We explicitly informed participants that both positive and negative feedback would be useful to the design team for making improvements.
A researcher took notes of verbal responses while examples were shown. % (not full transcriptions).
After seeing all examples, participants filled out a short survey asking what they thought performed well, needed improvement, and could fit into their design practice.
An entire session took approximately 45 minutes to complete.
For analysis, feedback from written responses was processed in an open coding phase, and further grouped by one researcher into the related topics, which were agreed upon with the other researchers~\cite{openCoding}. Quotes shared below are exclusively from survey responses.

\subsection{Results}

\subsubsection{Tappability predictions can save significant time and effort compared to user studies}
From survey responses, 11 participants perceived the system as accurate, and 7 remarked how the system would be valuable for evaluating designs as a time-saving alternative to running user studies:
\qtp{It's fairly accurate in predicting whether an element is tappable or not}{P7};
\qtp{I think it's great to see a quantified results of tappability - it can reduce the time to conduct usability study.}{P6};
\qtp{UI designers could use the model to cross check and see if they match the anticipated results. If that happens, it will save a lot of time running user studies.}{P3}.
One noteworthy theme was the value of using our system for rapid evaluations at multiple stages of the design process:
\qtp{It might be useful during handoff to engineers as a final check on design quality, or assessing a built app during a usability audit.}{P4};
\qtp{I would use it in the evaluative stages of design as a gut check on what I've done.}{P9}.
Our system could directly enable this capability if implemented in an end-to-end application, since it only requires pixels as input and thus can operate on fully implemented UIs or visually realistic mockups.
% ``The predictions are good for designer to see if they are on the right track'' (P10).

P5 pointed out a trade-off of using our system for evaluation, reflecting on its static nature, versus the open-ended format of traditional usability studies:
\qt{Users might have an advantage by being able to trial and error. It seems like the model gets a lot correct and points out possible design flaws, but users tend to explore openly anyway making choices still situational.}
Overall, this feedback suggests strong potential uses cases in rapid, heuristic evaluations of UIs when user studies would be too time-consuming, both for the early stages of design (when prototyping alternatives) and for catching potential errors in a design as it nears production.

\subsubsection{Analysis of a single screen offers limited notion of context in a UI flow}
While many participants remarked on the model's generally good performance, 4 were more critical or skeptical when it came to UI elements that were sensitive to the context of other screens in a UI flow, e.g.:
\qtp{It doesn't seem as useful for navigation or text where the tappability is more contextual}{P4};
\qtp{Considering context and looking at the whole page holistically are very important in UI design. The system tend to ignore the the context of the screen. E.g. Is it the home screen or interior page? The app logo can be tappable depending on the context}{P13}.
% A screenshot's text content and its place in a greater UI flow can have a significant impact on users' perceptions of tappability.
The cases referred to in these quotes are examples with low model confidence, reflecting a potentially ambiguous perception of tappability that depends on how the input screenshot is situated in a flow of multiple UIs.
One limitation of our model is that it only uses a \textit{static} snapshot of a UI as input. In future work, \textit{temporal} information could be used in our model's inputs to add additional context, as prior work has done for predicting user engagement with animations~\cite{wu_animations_2020} and grounding UI action sequences~\cite{li_grounding}. In addition, this is a case where including additional input modalities (e.g., text) can provide additional cues to boost prediction confidence.
% \qtp{Being able to search through similar examples would be useful to improve tappability or non-tappability depending on intent}{P2}

\subsubsection{Contrasting similar examples provided design feedback for iteration}
At least 5 participants wrote favorably of the contrasting similar examples in open-ended feedback, and remarked on their value for inspiring potential design changes:
\qtp{The initial "Model tap prob" metric is extremely useful as are the examples of similar UI elements that have both low and high [tappability] scores.}{P14};
\qtp{I might also use some of the comps [examples] to find inspirations on other ways to design a certain element}{P9}.
Some participants liked the diversity of some sets of contrasting similar examples
(\qtp{the provided examples are useful to reference and compare to, even if the similar elements are not exactly the same.}{P10}), while others desired a greater degree of semantic similarity
(\qtp{Heading component compares to a CTA button in Settings page. It feels like comparing apples to oranges. I would suggest, using similar UI component proximity for similar examples}{P1}).
One direction for future work could be to allow users to filter and set thresholds for examples (e.g., by certain types or locations of UI elements), or reporting actual distances (\qtp{Maybe for the nearest neighbors, provide some indications of how near or far the neighbor is, e.g. 90\% vs 10\%}{P12}).

Overall, this feedback suggests that the contrasting similar examples, curated based on a specified UI element, have the potential to provide useful inspiration for designers. This may help ``close the loop'' beyond tappability prediction scores alone.

\subsubsection{Participants desire more explicit explanations beyond the heatmap}
While 2 participants remarked that the heatmap was useful, 4 participants noted the heatmap was confusing to use, or needed better instructions, e.g.:
\qtp{Heat map. Confused me and would need some guidance on how to process the info.}{P14};
\qtp{Saliency heatmap definitely needs some mental shift to understand.}{P6}.
While this could potentially be mitigated with improved onboarding or more experience~\cite{cai_onboarding_2021, yang_uxml},
% the saliency heatmap may also be intrinsically limited in its application due to the ``black-box'' nature of our prediction model. This means, that while 
the ``black-box'' nature of our model means, while it may be effective at predicting users' perceptions of tappability, the mechanisms which enable those predictions may not reflect the same reasoning as users~\cite{lipton2017mythos}. The mismatch in mental models could explain this result:
\qtp{I found the heat maps and nearest neighbors less helpful because they didn't resemble my own mental model / instincts for evaluating the usability of these mockups}{P4}.

Some participants expressed a desire for deeper explanations of \textit{why} certain elements in the heatmap contribute to a tappability prediction more than others, which could help improve its usability:
\qtp{On the heat map, add some explanation about why the other elements might or might not impact the probability score of an elemement}{P14};
and others wished the system could output design suggestions directly:
\qtp{It'd be amazing if the system can provide recommendation like boost the color contrast}{P6}.
One promising direction for future work could draw from techniques in ML debugging research, by identifying common ``heuristics'' from patterns in the model's and XRAI heatmap's outputs and raising messages with concrete design suggestions (e.g., increasing contrast or changing colors)~\cite{umlaut, tutorons}.

% \subsubsection{Role of Prediction Model with Styleguides}
% \added{

% the system might not be as useful to designers who work mainly with styleguides, since these guides have best practices baked in already. However, for designers who make styleguides or operate in open ended design space, it can be very useful, and reinforce best practices.
% }

%% file: discussion.tex
\section{Themes in Model Behavior: Discussion and Limitations}
\label{sec:discussion}

In this section, we describe patterns observed in our selected examples and discuss implications for the use of our model and explanation mechanisms.

\subsection{Persistent Signals and Signifiers}

\paragraph{Text as a feature indicating tappability}
In examples in \autoref{sec:analysis} as well as examples in our user evaluation, bounding boxes surrounding text elements were highly attributed by XRAI. This does not necessarily mean the text itself is perceived as tappable, but rather that the existence of text serves as a signifier of tappability to nearby elements (see \autoref{sec:interpretability_challenges}).
This is one potential drawback to our pixel-based model compared to multimodal models that use text as input to gain a deeper understanding of an element's context (e.g., a ``submit'' button or ``click here to unsubscribe'' text).

\paragraph{Icons next to text generally indicate tappable regions}
To our model, small icons or graphics appearing next to text strongly signify tappability. This is demonstrated in \autoref{fig:stocks_textview_nns}, where most tappable text elements are near radio buttons, icons, and other graphics. Using icons to signify the tappability of adjacent text elements is a well-known practice~\cite{nn_affordances}.
However, our model does not always produce reliable tappability predictions of checkbox elements (E.g., \autoref{fig:stocks_textview_nns}, bottom right). This is likely due to ambiguity in the labeling task. Since the checkbox, accompanying description, and parent element containing both are each distinct UI elements with separate bounding boxes, any one of these elements within a given screenshot could be selected for labeling. Crowdworkers may have different perceptions of the tappability of the different elements, and this uncertainty is reflected in our model's prediction scores.

\paragraph{Image views in apps are not consistent predictors}
Because the content of \texttt{ImageView} elements can be highly varied (e.g., containing icons, logos, thumbnails, previews, ...), they can sometimes confound our purely vision-based model (see \autoref{fig:emergency_button_nns}). 
While our model likely also uses the location and context of the image element, the content of the image can overwhelm predictions, possibly due to the texture sensitivity of CNNs~\cite{cnn_texture}.
One way to potentially mitigate this effect would be to replace images with placeholders, similar to wireframes~\cite{deka_rico_2017, chen_wireframe-based_2020}.

\subsection{Challenges in Interpreting XRAI Attributions}
\label{sec:interpretability_challenges}

\paragraph{XRAI attributions highlight influential regions; highly influential regions are not necessarily tappable themselves}
As reflected in the results of our user evaluation, the XRAI heatmaps require practice to take full advantage of, and could benefit from the addition of heuristic-based explanations.
A critical note for our use of XRAI is that the heatmap it produces is not a tappability heatmap, but a heatmap showing how regions in the UI screen influence the tappability prediction \textit{for a particular element}.
For example, if highly attributed text near a button was removed, that button would likely no longer be classified as tappable.
As such, saliency methods like XRAI are often useful in practice for diagnosing the features that influence predictions, and the sensitivity of that prediction to contextual factors.

\paragraph{Summarizing attributions with regions may leave out important details}
In contrast to XRAI, which uses regions, pixel-based saliency methods like Integrated Gradients~\cite{ig_axiomatic} highlight inputs at a finer scale. While this may be useful for debugging features of small UI elements, pixel-based methods are known to be difficult to interpret by humans compared to region-based methods, and can be susceptible to errors~\cite{kapishnikov_xrai, been_moritz_saliency_checks}.

\paragraph{XRAI attribution values cannot be compared between examples}
Like other gradient-based saliency methods, the raw values of XRAI attributions are specific to input examples~\cite{been_moritz_saliency_checks}.
Some other algorithms, such as DeepSHAP~\cite{deepshap}, sum to the probabilities of predictions, and may be compared between examples. These other methods could also enable new interactions, such as aggregated analyses, a promising direction for future work.

\subsection{Browsing Nearest Neighbor Examples}
\label{sec:discussion_nns}

\paragraph{Nearest Neighbors capture many dimensions of similarity}
Across all of our examples, nearest neighbors appear to capture dimensions beyond visual similarity alone. In particular, bounding box locations, sizes, and aspect ratios are generally similar among neighbors. This indicates that our model has not only learned to use the appearance of an element to predict its tappability, but also contextual information such as its location, shape, and proximity to other elements. As participants in our user evaluation noted, adding interactivity to the nearest neighbor examples, such as the ability to filter and sort by component types and application properties, could help narrow down these contextual cues to provide more relevant feedback for iterating UI designs.

\paragraph{Splitting neighbors by binary tappability predictions discards some information}
While using a discrete boundary can provide useful contrasting examples, the average tappability prediction probabilities and distances between splits can contain subtle yet important information about the landscape of UI design patterns related to the input.
For example, the health app's non-tappable neighbors included many seemingly unrelated graphics. This may be because of skewed distances, i.e., most nearby examples are tappable, and the closest non-tappable neighors are significantly further away, and, thus, less similar.
In other cases, the sets of neighbors may have skewed average probabilities (e.g., near 99\% for tappable, and near 49\% for non tappable). This is an additional, strong, indicator that similar UIs are generally either perceived as tappable or uncertain, rather than non-tappable. The contrary is also true: the neighbors of confidently \textit{non-tappable} examples often score near 51\% for tappable, and 0\% for non-tappable.
In future work, these details could be made explicit in more continuous, interactive visualizations, to help designers explore related UI designs.

% \paragraph{Neither model predictions nor crowdworker labels are perfectly reliable. Consider votes}

\paragraph{Encouraging exploration of neighboring examples}
While using UIs with similar designs but different effects on tappability perception are useful for contrasting explanations, designers often value seeing diverse examples of designs for inspiration~\cite{chen_wireframe-based_2020, amanda_scout}. Future iterations of this work could sample more distant UI examples, filtered by UI element types, or even learned concepts, i.e., with concept vectors~\cite{tcav}.
In addition, since nearest neighbor examples are split by model predictions, the similar examples do not have to be limited to our source dataset. In other words, our model can be used to retrieve nearest neighbors or similar examples from other datasets.

\subsection{Additional Limitations}

\paragraph{Concept drift}
While UI design styles and trends change over time, our model is trained on a “static” snapshot of application UIs, and may give less reliable predictions over time. This phenomenon is known as concept drift~\cite{concept_drift}, and may be mitigated by augmenting the dataset with new examples over time.
Furthermore, since our dataset comprises only Android applications, our model may require fine-tuning to generalize well to UI screenshots from other platforms.

% \paragraph{Use of well-known UI libraries}
% Performance of the model on apps that instantiate best practices from UI libraries, compared to ad hoc designed apps is not well known

\paragraph{Perceived tappability predictions contain multiple signals}
As we have found by analyzing the distribution of label agreement in our dataset, the tappability of many UI elements in the wild appear ambiguous to users. While this uncertainty explicitly limits the possible accuracy of our model, it also means that predictions near the decision boundary suggest user confusion. This signal, along with other usability metrics (e.g., engagement~\cite{wu_animations_2020} or cognitive load) may be useful outputs from future models.

%% file: conclusion.tex
\section{Conclusion}
We presented a novel, automated system for predicting the human perceived tappability of mobile UI elements and explaining model predictions to users. Our work significantly advanced the art by developing a purely vision-based deep neural network, which only relies on pixels and does not require a UI to be fully specified; and by enabling mechanisms for explaining design insights to the user with contextual and instance-level interpretations of model predictions. We also create a new tappability dataset where each element is labeled by multiple crowdworkers for reliable tappability estimation. We provided an in-depth discussion of our model behavior and explanation mechanisms through extensive analysis of examples and collected feedback from experienced professional UI/UX designers in how they would use and improve our system. Together, our work not only advances tappability modeling research but also demonstrates how deep learning approaches can be used for automatic UI usability analysis.

%% file: study_UIs.tex
\section{User Study Figures}
\label{sec:user_study_figs}

Following are the UIs and outputs of our explanation algorithms shown to participants in the user evaluation in ~\autoref{sec:eval}.

\subsection{Example 1: Onboarding Example}
\centering\includegraphics[width=0.93\textwidth]{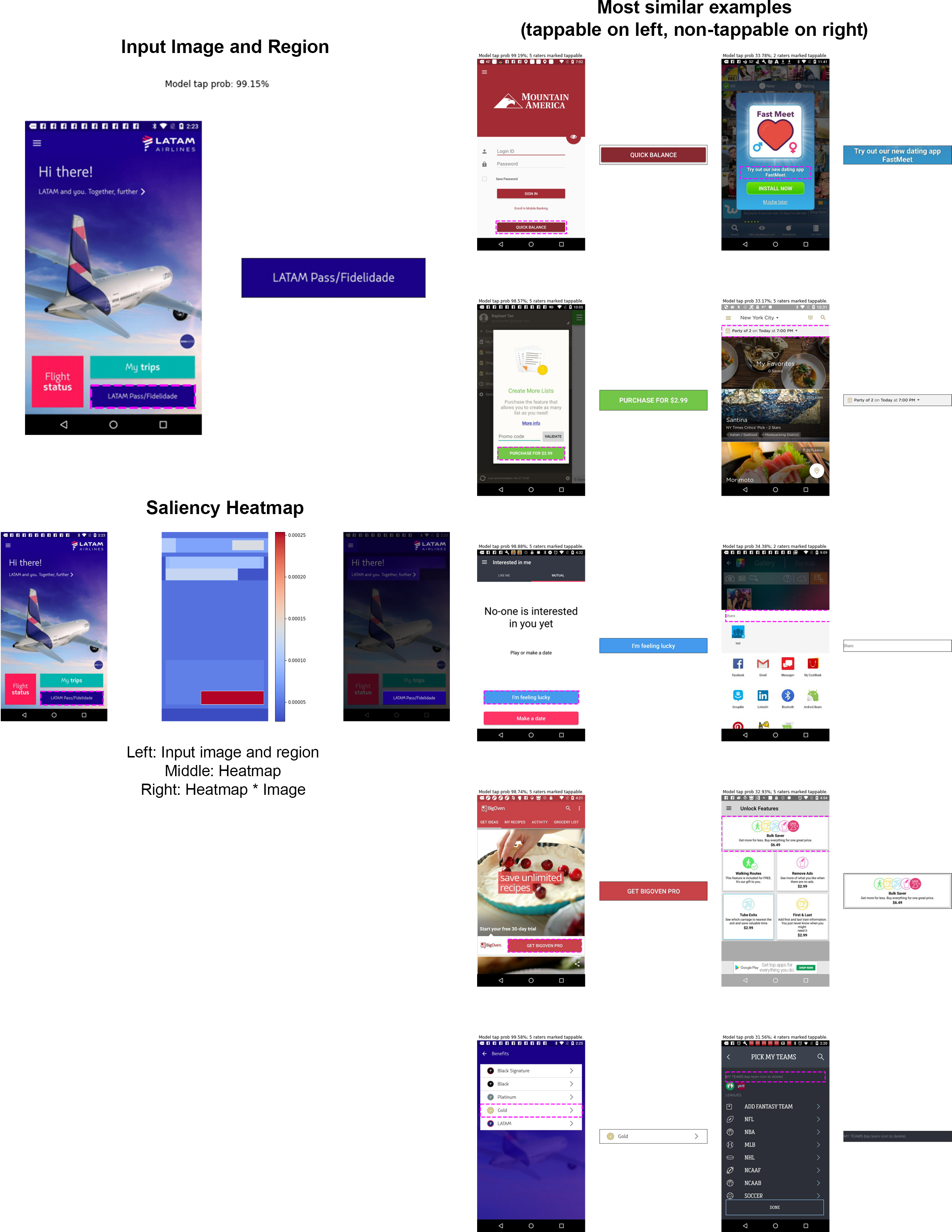}

\subsection{Example 2}

\centering\includegraphics[width=0.93\textwidth]{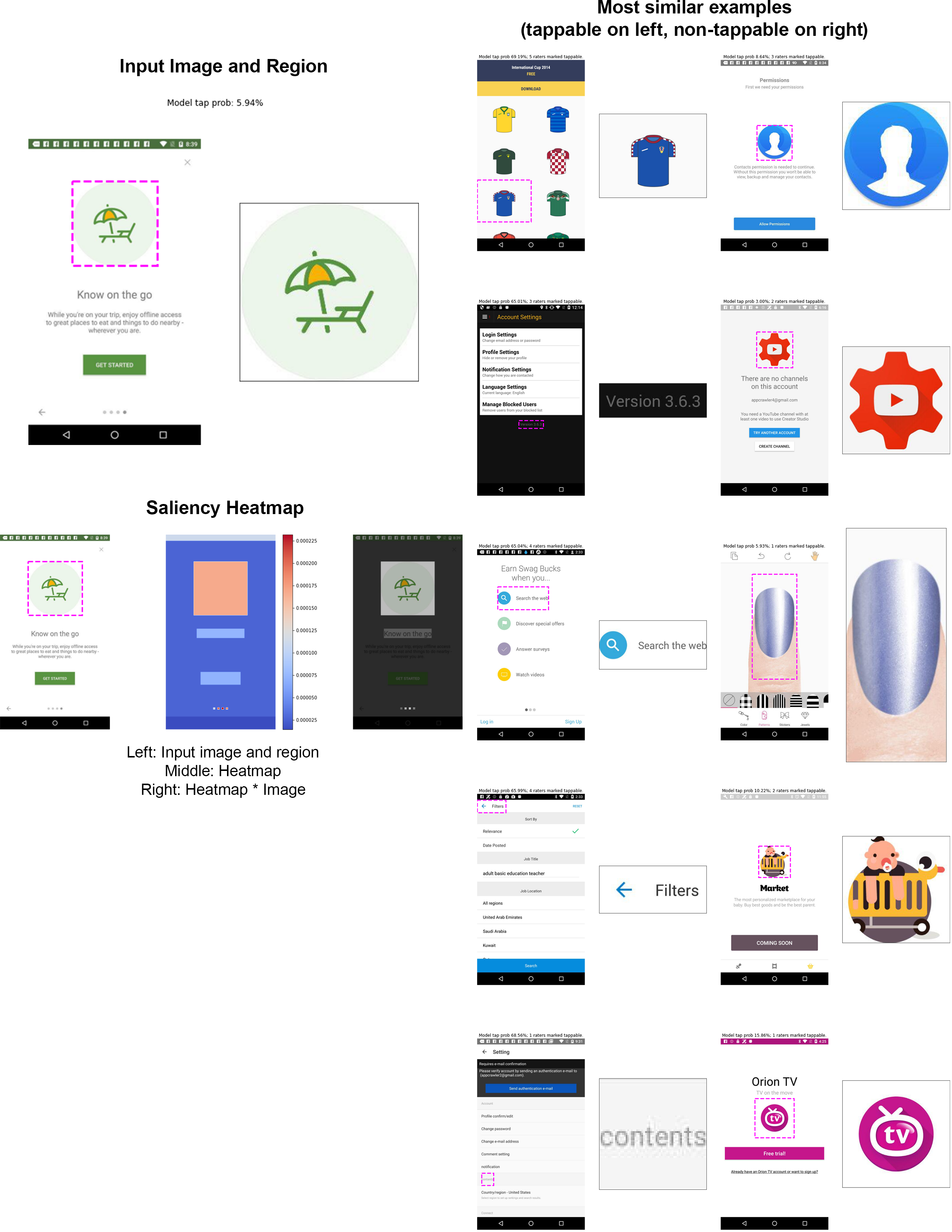}

\subsection{Example 3}

\centering\includegraphics[width=0.93\textwidth]{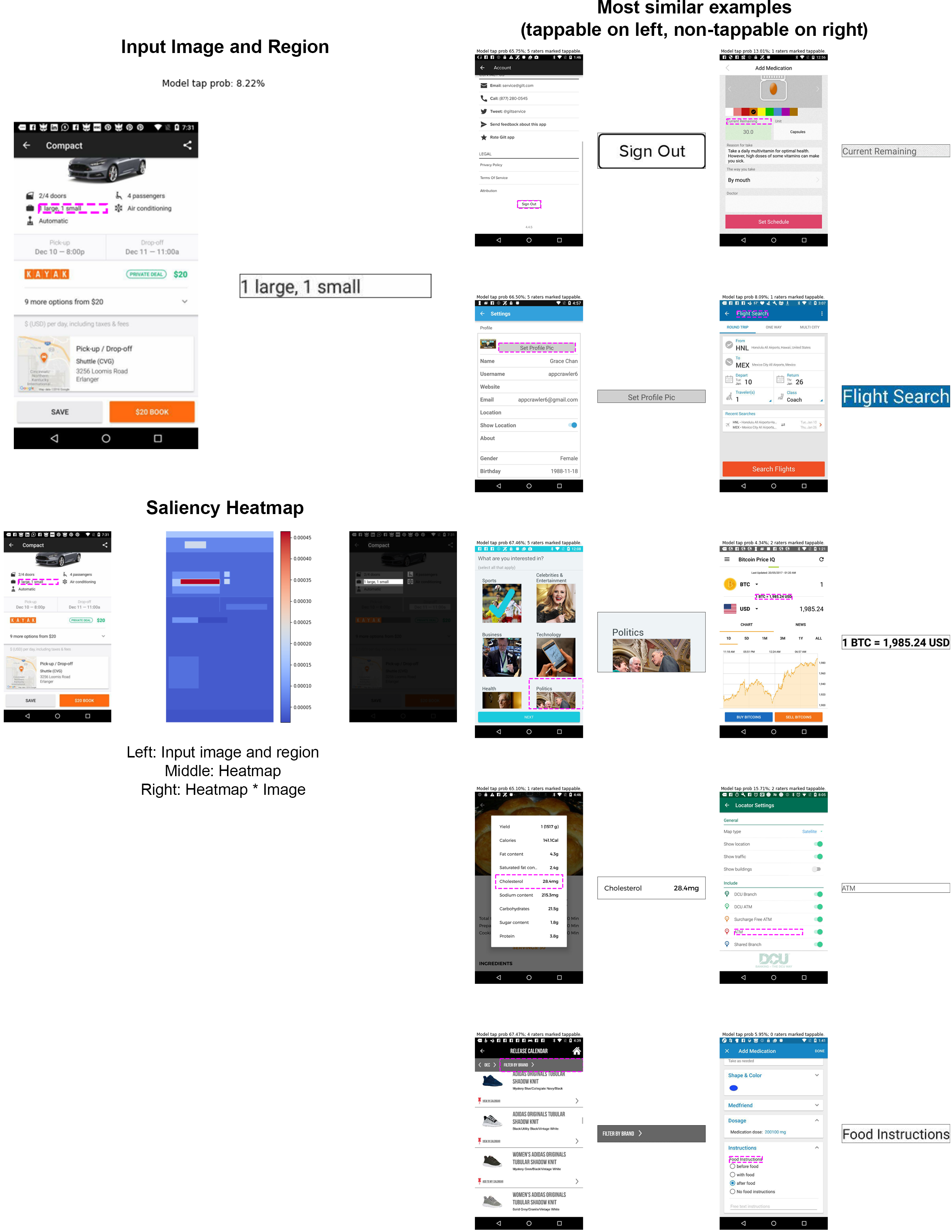}

\subsection{Example 4}

\centering\includegraphics[width=0.93\textwidth]{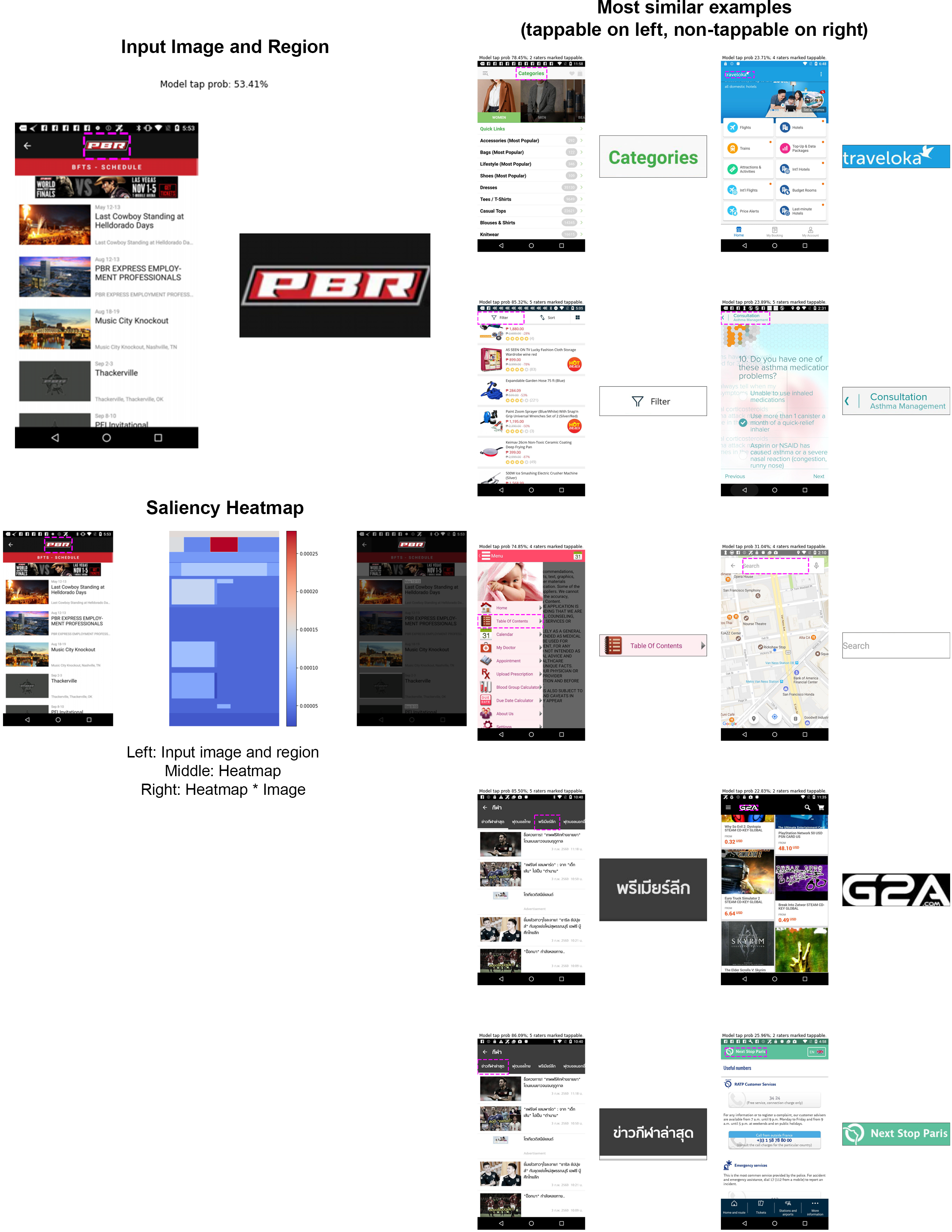}

\subsection{Example 5}

\centering\includegraphics[width=0.93\textwidth]{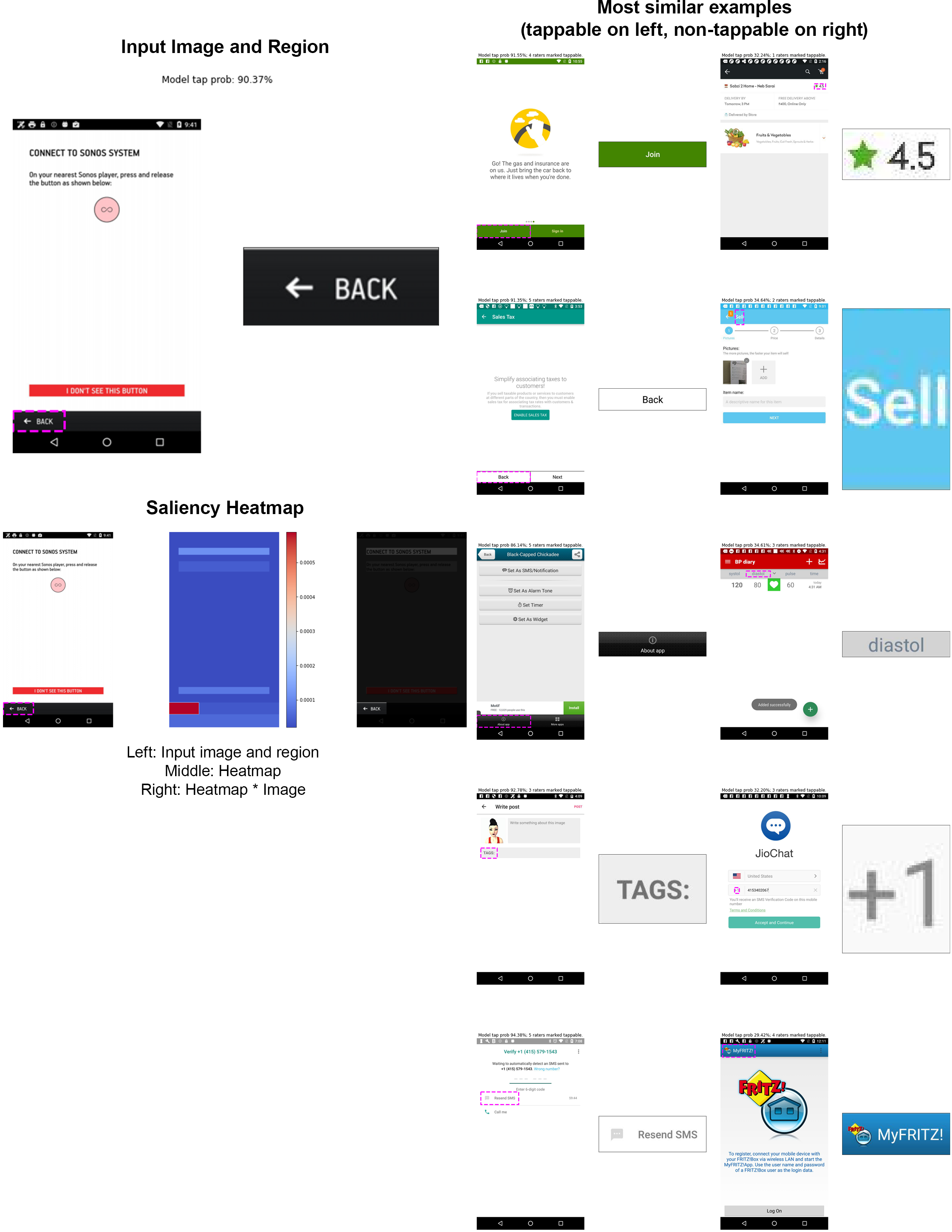}

\subsection{Example 6}

\centering\includegraphics[width=0.93\textwidth]{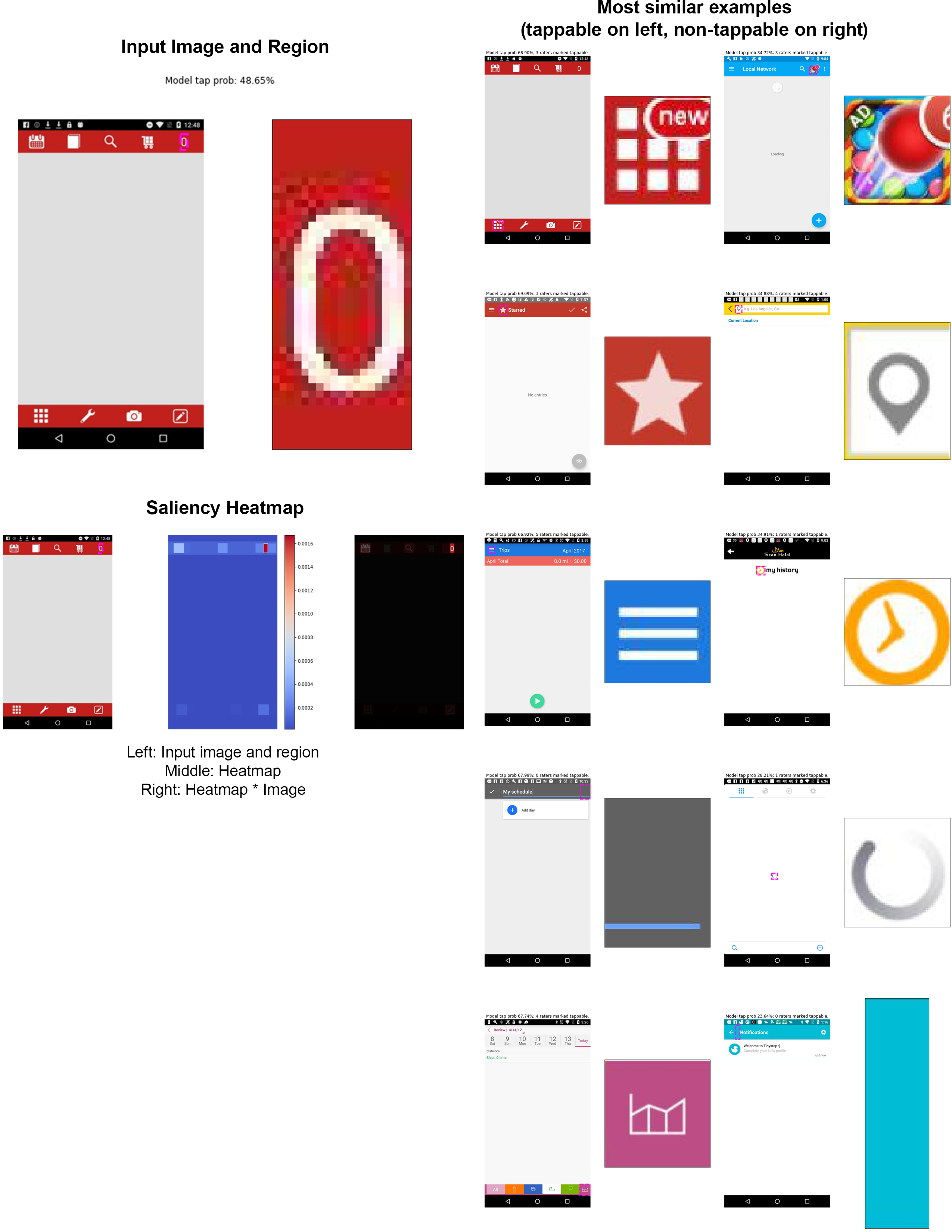}

%% file: main.bbl

\begin{thebibliography}{45}

%%% the Web address.

\ifx \showCODEN    \undefined \def \showCODEN     #1{\unskip}     \fi
\ifx \showDOI      \undefined \def \showDOI       #1{#1}\fi
\ifx \showISBNx    \undefined \def \showISBNx     #1{\unskip}     \fi
\ifx \showISBNxiii \undefined \def \showISBNxiii  #1{\unskip}     \fi
\ifx \showISSN     \undefined \def \showISSN      #1{\unskip}     \fi
\ifx \showLCCN     \undefined \def \showLCCN      #1{\unskip}     \fi
\ifx \shownote     \undefined \def \shownote      #1{#1}          \fi
\ifx \showarticletitle \undefined \def \showarticletitle #1{#1}   \fi
\ifx \showURL      \undefined \def \showURL       {\relax}        \fi
% invisible to TeX
\providecommand\bibfield[2]{#2}
\providecommand\bibinfo[2]{#2}
\providecommand\natexlab[1]{#1}
\providecommand\showeprint[2][]{arXiv:#2}

\bibitem[Adebayo et~al\mbox{.}(2018)]%
        {been_moritz_saliency_checks}
\bibfield{author}{\bibinfo{person}{Julius Adebayo}, \bibinfo{person}{Justin
  Gilmer}, \bibinfo{person}{Michael Muelly}, \bibinfo{person}{Ian Goodfellow},
  \bibinfo{person}{Moritz Hardt}, {and} \bibinfo{person}{Been Kim}.}
  \bibinfo{year}{2018}\natexlab{}.
\newblock \showarticletitle{Sanity Checks for Saliency Maps}. In
  \bibinfo{booktitle}{\emph{Advances in Neural Information Processing
  Systems}}, \bibfield{editor}{\bibinfo{person}{S.~Bengio},
  \bibinfo{person}{H.~Wallach}, \bibinfo{person}{H.~Larochelle},
  \bibinfo{person}{K.~Grauman}, \bibinfo{person}{N.~Cesa-Bianchi}, {and}
  \bibinfo{person}{R.~Garnett}} (Eds.), Vol.~\bibinfo{volume}{31}.
  \bibinfo{publisher}{Curran Associates, Inc.}
\newblock
\urldef\tempurl%
\url{https://proceedings.neurips.cc/paper/2018/file/294a8ed24b1ad22ec2e7efea049b8737-Paper.pdf}
\showURL{%
\tempurl}


\bibitem[Bai et~al\mbox{.}(2021)]%
        {UIBert}
\bibfield{author}{\bibinfo{person}{Chongyang Bai}, \bibinfo{person}{Xiaoxue
  Zang}, \bibinfo{person}{Ying Xu}, \bibinfo{person}{Srinivas Sunkara},
  \bibinfo{person}{Abhinav Rastogi}, \bibinfo{person}{Jindong Chen}, {and}
  \bibinfo{person}{Blaise~Ag{\"{u}}era y Arcas}.}
  \bibinfo{year}{2021}\natexlab{}.
\newblock \showarticletitle{UIBert: Learning Generic Multimodal Representations
  for {UI} Understanding}.
\newblock \bibinfo{journal}{\emph{CoRR}}  \bibinfo{volume}{abs/2107.13731}
  (\bibinfo{year}{2021}).
\newblock
\showeprint[arXiv]{2107.13731}
\urldef\tempurl%
\url{https://arxiv.org/abs/2107.13731}
\showURL{%
\tempurl}


\bibitem[Bellamy et~al\mbox{.}(2011)]%
        {cogtool}
\bibfield{author}{\bibinfo{person}{Rachel Bellamy}, \bibinfo{person}{Bonnie
  John}, {and} \bibinfo{person}{Sandra Kogan}.}
  \bibinfo{year}{2011}\natexlab{}.
\newblock \showarticletitle{Deploying CogTool: integrating quantitative
  usability assessment into real-world software development}. In
  \bibinfo{booktitle}{\emph{2011 33rd International Conference on Software
  Engineering (ICSE)}}. \bibinfo{pages}{691--700}.
\newblock
\urldef\tempurl%
\url{https://doi.org/10.1145/1985793.1985890}
\showDOI{\tempurl}


\bibitem[Bylinskii et~al\mbox{.}(2017)]%
        {bylinskii_learning_2017}
\bibfield{author}{\bibinfo{person}{Zoya Bylinskii}, \bibinfo{person}{Nam~Wook
  Kim}, \bibinfo{person}{Peter O'Donovan}, \bibinfo{person}{Sami Alsheikh},
  \bibinfo{person}{Spandan Madan}, \bibinfo{person}{Hanspeter Pfister},
  \bibinfo{person}{Fredo Durand}, \bibinfo{person}{Bryan Russell}, {and}
  \bibinfo{person}{Aaron Hertzmann}.} \bibinfo{year}{2017}\natexlab{}.
\newblock \showarticletitle{Learning {Visual} {Importance} for {Graphic}
  {Designs} and {Data} {Visualizations}}. In
  \bibinfo{booktitle}{\emph{Proceedings of the 30th {Annual} {ACM} {Symposium}
  on {User} {Interface} {Software} and {Technology}}}.
  \bibinfo{publisher}{ACM}, \bibinfo{address}{Québec City QC Canada},
  \bibinfo{pages}{57--69}.
\newblock
\showISBNx{978-1-4503-4981-9}
\urldef\tempurl%
\url{https://doi.org/10.1145/3126594.3126653}
\showDOI{\tempurl}


\bibitem[Cai et~al\mbox{.}(2019)]%
        {cai_contrasting}
\bibfield{author}{\bibinfo{person}{Carrie~J. Cai}, \bibinfo{person}{Jonas
  Jongejan}, {and} \bibinfo{person}{Jess Holbrook}.}
  \bibinfo{year}{2019}\natexlab{}.
\newblock \showarticletitle{The Effects of Example-Based Explanations in a
  Machine Learning Interface}. In \bibinfo{booktitle}{\emph{Proceedings of the
  24th International Conference on Intelligent User Interfaces}} (Marina del
  Ray, California) \emph{(\bibinfo{series}{IUI '19})}.
  \bibinfo{publisher}{Association for Computing Machinery},
  \bibinfo{address}{New York, NY, USA}, \bibinfo{pages}{258–262}.
\newblock
\showISBNx{9781450362726}
\urldef\tempurl%
\url{https://doi.org/10.1145/3301275.3302289}
\showDOI{\tempurl}


\bibitem[Cai et~al\mbox{.}(2021)]%
        {cai_onboarding_2021}
\bibfield{author}{\bibinfo{person}{Carrie~J Cai}, \bibinfo{person}{Samantha
  Winter}, \bibinfo{person}{David Steiner}, \bibinfo{person}{Lauren Wilcox},
  {and} \bibinfo{person}{Michael Terry}.} \bibinfo{year}{2021}\natexlab{}.
\newblock \showarticletitle{Onboarding {Materials} as {Cross}-functional
  {Boundary} {Objects} for {Developing} {AI} {Assistants}}. In
  \bibinfo{booktitle}{\emph{Extended {Abstracts} of the 2021 {CHI} {Conference}
  on {Human} {Factors} in {Computing} {Systems}}}. \bibinfo{publisher}{ACM},
  \bibinfo{address}{Yokohama Japan}, \bibinfo{pages}{1--7}.
\newblock
\showISBNx{978-1-4503-8095-9}
\urldef\tempurl%
\url{https://doi.org/10.1145/3411763.3443435}
\showDOI{\tempurl}


\bibitem[Chen et~al\mbox{.}(2019)]%
        {chen_gallery_2019}
\bibfield{author}{\bibinfo{person}{Chunyang Chen}, \bibinfo{person}{Sidong
  Feng}, \bibinfo{person}{Zhenchang Xing}, \bibinfo{person}{Linda Liu},
  \bibinfo{person}{Shengdong Zhao}, {and} \bibinfo{person}{Jinshui Wang}.}
  \bibinfo{year}{2019}\natexlab{}.
\newblock \showarticletitle{Gallery {D}.{C}.: {Design} {Search} and {Knowledge}
  {Discovery} through {Auto}-created {GUI} {Component} {Gallery}}.
\newblock \bibinfo{journal}{\emph{Proceedings of the ACM on Human-Computer
  Interaction}} \bibinfo{volume}{3}, \bibinfo{number}{CSCW}
  (\bibinfo{date}{Nov.} \bibinfo{year}{2019}), \bibinfo{pages}{1--22}.
\newblock
\showISSN{2573-0142}
\urldef\tempurl%
\url{https://doi.org/10.1145/3359282}
\showDOI{\tempurl}


\bibitem[Chen et~al\mbox{.}(2017)]%
        {chen_ui_2017}
\bibfield{author}{\bibinfo{person}{Chun-Fu~(Richard) Chen},
  \bibinfo{person}{Marco Pistoia}, \bibinfo{person}{Conglei Shi},
  \bibinfo{person}{Paolo Girolami}, \bibinfo{person}{Joseph~W. Ligman}, {and}
  \bibinfo{person}{Yong Wang}.} \bibinfo{year}{2017}\natexlab{}.
\newblock \showarticletitle{{UI} {X}-{Ray}: {Interactive} {Mobile} {UI}
  {Testing} {Based} on {Computer} {Vision}}. In
  \bibinfo{booktitle}{\emph{Proceedings of the 22nd {International}
  {Conference} on {Intelligent} {User} {Interfaces}}}.
  \bibinfo{publisher}{ACM}, \bibinfo{address}{Limassol Cyprus},
  \bibinfo{pages}{245--255}.
\newblock
\showISBNx{978-1-4503-4348-0}
\urldef\tempurl%
\url{https://doi.org/10.1145/3025171.3025190}
\showDOI{\tempurl}


\bibitem[Chen et~al\mbox{.}(2020)]%
        {chen_wireframe-based_2020}
\bibfield{author}{\bibinfo{person}{Jieshan Chen}, \bibinfo{person}{Chunyang
  Chen}, \bibinfo{person}{Zhenchang Xing}, \bibinfo{person}{Xin Xia},
  \bibinfo{person}{Liming Zhu}, \bibinfo{person}{John Grundy}, {and}
  \bibinfo{person}{Jinshui Wang}.} \bibinfo{year}{2020}\natexlab{}.
\newblock \showarticletitle{Wireframe-based {UI} {Design} {Search} through
  {Image} {Autoencoder}}.
\newblock \bibinfo{journal}{\emph{ACM Transactions on Software Engineering and
  Methodology}} \bibinfo{volume}{29}, \bibinfo{number}{3} (\bibinfo{date}{July}
  \bibinfo{year}{2020}), \bibinfo{pages}{1--31}.
\newblock
\showISSN{1049-331X, 1557-7392}
\urldef\tempurl%
\url{https://doi.org/10.1145/3391613}
\showDOI{\tempurl}


\bibitem[Cheng(2016)]%
        {variation_theory}
\bibfield{author}{\bibinfo{person}{W.~L. Cheng}.}
  \bibinfo{year}{2016}\natexlab{}.
\newblock \showarticletitle{Learning through the variation theory: A case
  study}.
\newblock \bibinfo{journal}{\emph{The International Journal of Teaching and
  Learning in Higher Education}}  \bibinfo{volume}{28} (\bibinfo{year}{2016}),
  \bibinfo{pages}{283--292}.
\newblock


\bibitem[Deka et~al\mbox{.}(2017a)]%
        {deka_rico_2017}
\bibfield{author}{\bibinfo{person}{Biplab Deka}, \bibinfo{person}{Zifeng
  Huang}, \bibinfo{person}{Chad Franzen}, \bibinfo{person}{Joshua Hibschman},
  \bibinfo{person}{Daniel Afergan}, \bibinfo{person}{Yang Li},
  \bibinfo{person}{Jeffrey Nichols}, {and} \bibinfo{person}{Ranjitha Kumar}.}
  \bibinfo{year}{2017}\natexlab{a}.
\newblock \showarticletitle{Rico: {A} {Mobile} {App} {Dataset} for {Building}
  {Data}-{Driven} {Design} {Applications}}. In
  \bibinfo{booktitle}{\emph{Proceedings of the 30th {Annual} {ACM} {Symposium}
  on {User} {Interface} {Software} and {Technology}}}.
  \bibinfo{publisher}{ACM}, \bibinfo{address}{Québec City QC Canada},
  \bibinfo{pages}{845--854}.
\newblock
\showISBNx{978-1-4503-4981-9}
\urldef\tempurl%
\url{https://doi.org/10.1145/3126594.3126651}
\showDOI{\tempurl}


\bibitem[Deka et~al\mbox{.}(2017b)]%
        {zipt}
\bibfield{author}{\bibinfo{person}{Biplab Deka}, \bibinfo{person}{Zifeng
  Huang}, \bibinfo{person}{Chad Franzen}, \bibinfo{person}{Jeffrey Nichols},
  \bibinfo{person}{Yang Li}, {and} \bibinfo{person}{Ranjitha Kumar}.}
  \bibinfo{year}{2017}\natexlab{b}.
\newblock \showarticletitle{ZIPT: Zero-Integration Performance Testing of
  Mobile App Designs}. In \bibinfo{booktitle}{\emph{Proceedings of the 30th
  Annual ACM Symposium on User Interface Software and Technology}} (Qu\'{e}bec
  City, QC, Canada) \emph{(\bibinfo{series}{UIST '17})}.
  \bibinfo{publisher}{Association for Computing Machinery},
  \bibinfo{address}{New York, NY, USA}, \bibinfo{pages}{727–736}.
\newblock
\showISBNx{9781450349819}
\urldef\tempurl%
\url{https://doi.org/10.1145/3126594.3126647}
\showDOI{\tempurl}


\bibitem[Deka et~al\mbox{.}(2016)]%
        {erica}
\bibfield{author}{\bibinfo{person}{Biplab Deka}, \bibinfo{person}{Zifeng
  Huang}, {and} \bibinfo{person}{Ranjitha Kumar}.}
  \bibinfo{year}{2016}\natexlab{}.
\newblock \showarticletitle{ERICA: Interaction Mining Mobile Apps}. In
  \bibinfo{booktitle}{\emph{Proceedings of the 29th Annual Symposium on User
  Interface Software and Technology}} (Tokyo, Japan)
  \emph{(\bibinfo{series}{UIST '16})}. \bibinfo{publisher}{Association for
  Computing Machinery}, \bibinfo{address}{New York, NY, USA},
  \bibinfo{pages}{767–776}.
\newblock
\showISBNx{9781450341899}
\urldef\tempurl%
\url{https://doi.org/10.1145/2984511.2984581}
\showDOI{\tempurl}


\bibitem[Fosco et~al\mbox{.}(2020)]%
        {predicting_visual_importance}
\bibfield{author}{\bibinfo{person}{Camilo Fosco}, \bibinfo{person}{Vincent
  Casser}, \bibinfo{person}{Amish~Kumar Bedi}, \bibinfo{person}{Peter
  O'Donovan}, \bibinfo{person}{Aaron Hertzmann}, {and} \bibinfo{person}{Zoya
  Bylinskii}.} \bibinfo{year}{2020}\natexlab{}.
\newblock \showarticletitle{Predicting Visual Importance Across Graphic Design
  Types}. In \bibinfo{booktitle}{\emph{Proceedings of the 33rd Annual ACM
  Symposium on User Interface Software and Technology}} (Virtual Event, USA)
  \emph{(\bibinfo{series}{UIST '20})}. \bibinfo{publisher}{Association for
  Computing Machinery}, \bibinfo{address}{New York, NY, USA},
  \bibinfo{pages}{249–260}.
\newblock
\showISBNx{9781450375146}
\urldef\tempurl%
\url{https://doi.org/10.1145/3379337.3415825}
\showDOI{\tempurl}


\bibitem[Geirhos et~al\mbox{.}(2018)]%
        {cnn_texture}
\bibfield{author}{\bibinfo{person}{Robert Geirhos}, \bibinfo{person}{Patricia
  Rubisch}, \bibinfo{person}{Claudio Michaelis}, \bibinfo{person}{Matthias
  Bethge}, \bibinfo{person}{Felix~A. Wichmann}, {and} \bibinfo{person}{Wieland
  Brendel}.} \bibinfo{year}{2018}\natexlab{}.
\newblock \showarticletitle{ImageNet-trained CNNs are biased towards texture;
  increasing shape bias improves accuracy and robustness}.
\newblock \bibinfo{journal}{\emph{CoRR}}  \bibinfo{volume}{abs/1811.12231}
  (\bibinfo{year}{2018}).
\newblock
\showeprint[arXiv]{1811.12231}
\urldef\tempurl%
\url{http://arxiv.org/abs/1811.12231}
\showURL{%
\tempurl}


\bibitem[Ghorbani et~al\mbox{.}(2019)]%
        {been_ace}
\bibfield{author}{\bibinfo{person}{Amirata Ghorbani}, \bibinfo{person}{James
  Wexler}, \bibinfo{person}{James~Y Zou}, {and} \bibinfo{person}{Been Kim}.}
  \bibinfo{year}{2019}\natexlab{}.
\newblock \showarticletitle{Towards Automatic Concept-based Explanations}. In
  \bibinfo{booktitle}{\emph{Advances in Neural Information Processing
  Systems}}, \bibfield{editor}{\bibinfo{person}{H.~Wallach},
  \bibinfo{person}{H.~Larochelle}, \bibinfo{person}{A.~Beygelzimer},
  \bibinfo{person}{F.~d\textquotesingle Alch\'{e}-Buc},
  \bibinfo{person}{E.~Fox}, {and} \bibinfo{person}{R.~Garnett}} (Eds.),
  Vol.~\bibinfo{volume}{32}. \bibinfo{publisher}{Curran Associates, Inc.}
\newblock
\urldef\tempurl%
\url{https://proceedings.neurips.cc/paper/2019/file/77d2afcb31f6493e350fca61764efb9a-Paper.pdf}
\showURL{%
\tempurl}


\bibitem[He et~al\mbox{.}(2016)]%
        {resnet_he}
\bibfield{author}{\bibinfo{person}{Kaiming He}, \bibinfo{person}{Xiangyu
  Zhang}, \bibinfo{person}{Shaoqing Ren}, {and} \bibinfo{person}{Jian Sun}.}
  \bibinfo{year}{2016}\natexlab{}.
\newblock \showarticletitle{Deep Residual Learning for Image Recognition}. In
  \bibinfo{booktitle}{\emph{2016 IEEE Conference on Computer Vision and Pattern
  Recognition (CVPR)}}. \bibinfo{pages}{770--778}.
\newblock
\urldef\tempurl%
\url{https://doi.org/10.1109/CVPR.2016.90}
\showDOI{\tempurl}


\bibitem[He et~al\mbox{.}(2020)]%
        {ActionBERT}
\bibfield{author}{\bibinfo{person}{Zecheng He}, \bibinfo{person}{Srinivas
  Sunkara}, \bibinfo{person}{Xiaoxue Zang}, \bibinfo{person}{Ying Xu},
  \bibinfo{person}{Lijuan Liu}, \bibinfo{person}{Nevan Wichers},
  \bibinfo{person}{Gabriel Schubiner}, \bibinfo{person}{Ruby~B. Lee}, {and}
  \bibinfo{person}{Jindong Chen}.} \bibinfo{year}{2020}\natexlab{}.
\newblock \showarticletitle{ActionBert: Leveraging User Actions for Semantic
  Understanding of User Interfaces}.
\newblock \bibinfo{journal}{\emph{CoRR}}  \bibinfo{volume}{abs/2012.12350}
  (\bibinfo{year}{2020}).
\newblock
\showeprint[arXiv]{2012.12350}
\urldef\tempurl%
\url{https://arxiv.org/abs/2012.12350}
\showURL{%
\tempurl}


\bibitem[Head et~al\mbox{.}(2015)]%
        {tutorons}
\bibfield{author}{\bibinfo{person}{Andrew Head}, \bibinfo{person}{Codanda
  Appachu}, \bibinfo{person}{Marti~A. Hearst}, {and} \bibinfo{person}{Björn
  Hartmann}.} \bibinfo{year}{2015}\natexlab{}.
\newblock \showarticletitle{Tutorons: Generating context-relevant, on-demand
  explanations and demonstrations of online code}. In
  \bibinfo{booktitle}{\emph{2015 IEEE Symposium on Visual Languages and
  Human-Centric Computing (VL/HCC)}}. \bibinfo{pages}{3--12}.
\newblock
\urldef\tempurl%
\url{https://doi.org/10.1109/VLHCC.2015.7356972}
\showDOI{\tempurl}


\bibitem[Huang et~al\mbox{.}(2019)]%
        {swire}
\bibfield{author}{\bibinfo{person}{Forrest Huang}, \bibinfo{person}{John~F.
  Canny}, {and} \bibinfo{person}{Jeffrey Nichols}.}
  \bibinfo{year}{2019}\natexlab{}.
\newblock \bibinfo{booktitle}{\emph{Swire: Sketch-Based User Interface
  Retrieval}}.
\newblock \bibinfo{publisher}{Association for Computing Machinery},
  \bibinfo{address}{New York, NY, USA}, \bibinfo{pages}{1–10}.
\newblock
\showISBNx{9781450359702}
\urldef\tempurl%
\url{https://doi.org/10.1145/3290605.3300334}
\showURL{%
\tempurl}


\bibitem[Kapishnikov et~al\mbox{.}(2019)]%
        {kapishnikov_xrai}
\bibfield{author}{\bibinfo{person}{Andrei Kapishnikov}, \bibinfo{person}{Tolga
  Bolukbasi}, \bibinfo{person}{Fernanda Viegas}, {and} \bibinfo{person}{Michael
  Terry}.} \bibinfo{year}{2019}\natexlab{}.
\newblock \showarticletitle{XRAI: Better Attributions Through Regions}. In
  \bibinfo{booktitle}{\emph{2019 IEEE/CVF International Conference on Computer
  Vision (ICCV)}}. \bibinfo{publisher}{IEEE}, \bibinfo{pages}{4947–4956}.
\newblock
\showISBNx{978-1-72814-803-8}
\urldef\tempurl%
\url{https://doi.org/10.1109/ICCV.2019.00505}
\showDOI{\tempurl}


\bibitem[Kim et~al\mbox{.}(2018)]%
        {tcav}
\bibfield{author}{\bibinfo{person}{Been Kim}, \bibinfo{person}{Martin
  Wattenberg}, \bibinfo{person}{Justin Gilmer}, \bibinfo{person}{Carrie~J.
  Cai}, \bibinfo{person}{James Wexler}, \bibinfo{person}{Fernanda~B.
  Vi{\'{e}}gas}, {and} \bibinfo{person}{Rory Sayres}.}
  \bibinfo{year}{2018}\natexlab{}.
\newblock \showarticletitle{Interpretability Beyond Feature Attribution:
  Quantitative Testing with Concept Activation Vectors {(TCAV)}}. In
  \bibinfo{booktitle}{\emph{Proceedings of the 35th International Conference on
  Machine Learning, {ICML} 2018, Stockholmsm{\"{a}}ssan, Stockholm, Sweden,
  July 10-15, 2018}} \emph{(\bibinfo{series}{Proceedings of Machine Learning
  Research}, Vol.~\bibinfo{volume}{80})},
  \bibfield{editor}{\bibinfo{person}{Jennifer~G. Dy} {and}
  \bibinfo{person}{Andreas Krause}} (Eds.). \bibinfo{publisher}{{PMLR}},
  \bibinfo{pages}{2673--2682}.
\newblock
\urldef\tempurl%
\url{http://proceedings.mlr.press/v80/kim18d.html}
\showURL{%
\tempurl}


\bibitem[Lee et~al\mbox{.}(2020)]%
        {guicomp}
\bibfield{author}{\bibinfo{person}{Chunggi Lee}, \bibinfo{person}{Sanghoon
  Kim}, \bibinfo{person}{Dongyun Han}, \bibinfo{person}{Hongjun Yang},
  \bibinfo{person}{Young-Woo Park}, \bibinfo{person}{Bum~Chul Kwon}, {and}
  \bibinfo{person}{Sungahn Ko}.} \bibinfo{year}{2020}\natexlab{}.
\newblock \bibinfo{booktitle}{\emph{GUIComp: A GUI Design Assistant with
  Real-Time, Multi-Faceted Feedback}}.
\newblock \bibinfo{publisher}{Association for Computing Machinery},
  \bibinfo{address}{New York, NY, USA}, \bibinfo{pages}{1–13}.
\newblock
\showISBNx{9781450367080}
\urldef\tempurl%
\url{https://doi.org/10.1145/3313831.3376327}
\showURL{%
\tempurl}


\bibitem[Leiva et~al\mbox{.}(2020)]%
        {enrico}
\bibfield{author}{\bibinfo{person}{Luis~A. Leiva}, \bibinfo{person}{Asutosh
  Hota}, {and} \bibinfo{person}{Antti Oulasvirta}.}
  \bibinfo{year}{2020}\natexlab{}.
\newblock \showarticletitle{Enrico: A Dataset for Topic Modeling of Mobile UI
  Designs}. In \bibinfo{booktitle}{\emph{22nd International Conference on
  Human-Computer Interaction with Mobile Devices and Services}} (Oldenburg,
  Germany) \emph{(\bibinfo{series}{MobileHCI '20})}.
  \bibinfo{publisher}{Association for Computing Machinery},
  \bibinfo{address}{New York, NY, USA}, Article \bibinfo{articleno}{9},
  \bibinfo{numpages}{4}~pages.
\newblock
\showISBNx{9781450380522}
\urldef\tempurl%
\url{https://doi.org/10.1145/3406324.3410710}
\showDOI{\tempurl}


\bibitem[Li et~al\mbox{.}(2021)]%
        {screen2vec}
\bibfield{author}{\bibinfo{person}{Toby Jia-Jun Li}, \bibinfo{person}{Lindsay
  Popowski}, \bibinfo{person}{Tom Mitchell}, {and} \bibinfo{person}{Brad~A
  Myers}.} \bibinfo{year}{2021}\natexlab{}.
\newblock \bibinfo{booktitle}{\emph{Screen2Vec: Semantic Embedding of GUI
  Screens and GUI Components}}.
\newblock \bibinfo{publisher}{Association for Computing Machinery},
  \bibinfo{address}{New York, NY, USA}.
\newblock
\showISBNx{9781450380966}
\urldef\tempurl%
\url{https://doi.org/10.1145/3411764.3445049}
\showURL{%
\tempurl}


\bibitem[Li et~al\mbox{.}(2020a)]%
        {li_grounding}
\bibfield{author}{\bibinfo{person}{Yang Li}, \bibinfo{person}{Jiacong He},
  \bibinfo{person}{Xin Zhou}, \bibinfo{person}{Yuan Zhang}, {and}
  \bibinfo{person}{Jason Baldridge}.} \bibinfo{year}{2020}\natexlab{a}.
\newblock \showarticletitle{Mapping Natural Language Instructions to Mobile UI
  Action Sequences}. In \bibinfo{booktitle}{\emph{Annual Conference of the
  Association for Computational Linguistics (ACL 2020)}}.
\newblock
\urldef\tempurl%
\url{https://www.aclweb.org/anthology/2020.acl-main.729.pdf}
\showURL{%
\tempurl}


\bibitem[Li et~al\mbox{.}(2020b)]%
        {ai_hci_workshop}
\bibfield{author}{\bibinfo{person}{Yang Li}, \bibinfo{person}{Ranjitha Kumar},
  \bibinfo{person}{Walter~S. Lasecki}, {and} \bibinfo{person}{Otmar Hilliges}.}
  \bibinfo{year}{2020}\natexlab{b}.
\newblock \showarticletitle{Artificial Intelligence for HCI: A Modern
  Approach}. In \bibinfo{booktitle}{\emph{Extended Abstracts of the 2020 CHI
  Conference on Human Factors in Computing Systems}} (Honolulu, HI, USA)
  \emph{(\bibinfo{series}{CHI EA '20})}. \bibinfo{publisher}{Association for
  Computing Machinery}, \bibinfo{address}{New York, NY, USA},
  \bibinfo{pages}{1–8}.
\newblock
\showISBNx{9781450368193}
\urldef\tempurl%
\url{https://doi.org/10.1145/3334480.3375147}
\showDOI{\tempurl}


\bibitem[Lipton(2017)]%
        {lipton2017mythos}
\bibfield{author}{\bibinfo{person}{Zachary~C. Lipton}.}
  \bibinfo{year}{2017}\natexlab{}.
\newblock \bibinfo{title}{The Mythos of Model Interpretability}.
\newblock
\newblock
\showeprint[arxiv]{1606.03490}~[cs.LG]


\bibitem[Loranger(2015)]%
        {nn_affordances}
\bibfield{author}{\bibinfo{person}{Hoa Loranger}.}
  \bibinfo{year}{2015}\natexlab{}.
\newblock \bibinfo{booktitle}{\emph{Beyond Blue Links: Making Clickable
  Elements Recognizable}}.
\newblock Nielsen Norman Group.
\newblock
\urldef\tempurl%
\url{https://www.nngroup.com/articles/clickable-elements/}
\showURL{%
\tempurl}


\bibitem[Lundberg and Lee(2017)]%
        {deepshap}
\bibfield{author}{\bibinfo{person}{Scott~M. Lundberg} {and}
  \bibinfo{person}{Su-In Lee}.} \bibinfo{year}{2017}\natexlab{}.
\newblock \showarticletitle{A Unified Approach to Interpreting Model
  Predictions}. In \bibinfo{booktitle}{\emph{Proceedings of the 31st
  International Conference on Neural Information Processing Systems}} (Long
  Beach, California, USA) \emph{(\bibinfo{series}{NIPS'17})}.
  \bibinfo{publisher}{Curran Associates Inc.}, \bibinfo{address}{Red Hook, NY,
  USA}, \bibinfo{pages}{4768–4777}.
\newblock
\showISBNx{9781510860964}


\bibitem[Papernot and McDaniel(2018)]%
        {deep_knn}
\bibfield{author}{\bibinfo{person}{Nicolas Papernot} {and}
  \bibinfo{person}{Patrick~D. McDaniel}.} \bibinfo{year}{2018}\natexlab{}.
\newblock \showarticletitle{Deep k-Nearest Neighbors: Towards Confident,
  Interpretable and Robust Deep Learning}.
\newblock \bibinfo{journal}{\emph{CoRR}}  \bibinfo{volume}{abs/1803.04765}
  (\bibinfo{year}{2018}).
\newblock
\showeprint[arXiv]{1803.04765}
\urldef\tempurl%
\url{http://arxiv.org/abs/1803.04765}
\showURL{%
\tempurl}


\bibitem[Ribeiro et~al\mbox{.}(2016)]%
        {lime}
\bibfield{author}{\bibinfo{person}{Marco~Tulio Ribeiro},
  \bibinfo{person}{Sameer Singh}, {and} \bibinfo{person}{Carlos Guestrin}.}
  \bibinfo{year}{2016}\natexlab{}.
\newblock \showarticletitle{"Why Should I Trust You?": Explaining the
  Predictions of Any Classifier}. In \bibinfo{booktitle}{\emph{Proceedings of
  the 22nd ACM SIGKDD International Conference on Knowledge Discovery and Data
  Mining}} (San Francisco, California, USA) \emph{(\bibinfo{series}{KDD '16})}.
  \bibinfo{publisher}{Association for Computing Machinery},
  \bibinfo{address}{New York, NY, USA}, \bibinfo{pages}{1135–1144}.
\newblock
\showISBNx{9781450342322}
\urldef\tempurl%
\url{https://doi.org/10.1145/2939672.2939778}
\showDOI{\tempurl}


\bibitem[Schoop et~al\mbox{.}(2021)]%
        {umlaut}
\bibfield{author}{\bibinfo{person}{Eldon Schoop}, \bibinfo{person}{Forrest
  Huang}, {and} \bibinfo{person}{Bjoern Hartmann}.}
  \bibinfo{year}{2021}\natexlab{}.
\newblock \bibinfo{booktitle}{\emph{UMLAUT: Debugging Deep Learning Programs
  Using Program Structure and Model Behavior}}.
\newblock \bibinfo{publisher}{Association for Computing Machinery},
  \bibinfo{address}{New York, NY, USA}.
\newblock
\showISBNx{9781450380966}
\urldef\tempurl%
\url{https://doi.org/10.1145/3411764.3445538}
\showURL{%
\tempurl}


\bibitem[Schrouff et~al\mbox{.}(2021)]%
        {local_global}
\bibfield{author}{\bibinfo{person}{Jessica Schrouff},
  \bibinfo{person}{Sebastien Baur}, \bibinfo{person}{Shaobo Hou},
  \bibinfo{person}{Diana Mincu}, \bibinfo{person}{Eric Loreaux},
  \bibinfo{person}{Ralph Blanes}, \bibinfo{person}{James Wexler},
  \bibinfo{person}{Alan Karthikesalingam}, {and} \bibinfo{person}{Been Kim}.}
  \bibinfo{year}{2021}\natexlab{}.
\newblock \showarticletitle{Best of both worlds: local and global explanations
  with human-understandable concepts}.
\newblock \bibinfo{journal}{\emph{CoRR}}  \bibinfo{volume}{abs/2106.08641}
  (\bibinfo{year}{2021}).
\newblock
\showeprint[arXiv]{2106.08641}
\urldef\tempurl%
\url{https://arxiv.org/abs/2106.08641}
\showURL{%
\tempurl}


\bibitem[Selvaraju et~al\mbox{.}(2017)]%
        {gradcam}
\bibfield{author}{\bibinfo{person}{Ramprasaath~R. Selvaraju},
  \bibinfo{person}{Michael Cogswell}, \bibinfo{person}{Abhishek Das},
  \bibinfo{person}{Ramakrishna Vedantam}, \bibinfo{person}{Devi Parikh}, {and}
  \bibinfo{person}{Dhruv Batra}.} \bibinfo{year}{2017}\natexlab{}.
\newblock \showarticletitle{Grad-CAM: Visual Explanations from Deep Networks
  via Gradient-Based Localization}. In \bibinfo{booktitle}{\emph{2017 IEEE
  International Conference on Computer Vision (ICCV)}}.
  \bibinfo{pages}{618--626}.
\newblock
\urldef\tempurl%
\url{https://doi.org/10.1109/ICCV.2017.74}
\showDOI{\tempurl}


\bibitem[Sundararajan et~al\mbox{.}(2017)]%
        {ig_axiomatic}
\bibfield{author}{\bibinfo{person}{Mukund Sundararajan}, \bibinfo{person}{Ankur
  Taly}, {and} \bibinfo{person}{Qiqi Yan}.} \bibinfo{year}{2017}\natexlab{}.
\newblock \showarticletitle{Axiomatic Attribution for Deep Networks}. In
  \bibinfo{booktitle}{\emph{Proceedings of the 34th International Conference on
  Machine Learning - Volume 70}} (Sydney, NSW, Australia)
  \emph{(\bibinfo{series}{ICML'17})}. \bibinfo{publisher}{JMLR.org},
  \bibinfo{pages}{3319–3328}.
\newblock


\bibitem[Swearngin and Li(2019)]%
        {swearngin_modeling_2019}
\bibfield{author}{\bibinfo{person}{Amanda Swearngin} {and}
  \bibinfo{person}{Yang Li}.} \bibinfo{year}{2019}\natexlab{}.
\newblock \showarticletitle{Modeling {Mobile} {Interface} {Tappability} {Using}
  {Crowdsourcing} and {Deep} {Learning}}. In
  \bibinfo{booktitle}{\emph{Proceedings of the 2019 {CHI} {Conference} on
  {Human} {Factors} in {Computing} {Systems}}}. \bibinfo{publisher}{ACM},
  \bibinfo{address}{Glasgow Scotland Uk}, \bibinfo{pages}{1--11}.
\newblock
\showISBNx{978-1-4503-5970-2}
\urldef\tempurl%
\url{https://doi.org/10.1145/3290605.3300305}
\showDOI{\tempurl}


\bibitem[Swearngin et~al\mbox{.}(2020)]%
        {amanda_scout}
\bibfield{author}{\bibinfo{person}{Amanda Swearngin},
  \bibinfo{person}{Chenglong Wang}, \bibinfo{person}{Alannah Oleson},
  \bibinfo{person}{James Fogarty}, {and} \bibinfo{person}{Amy~J. Ko}.}
  \bibinfo{year}{2020}\natexlab{}.
\newblock \bibinfo{booktitle}{\emph{Scout: Rapid Exploration of Interface
  Layout Alternatives through High-Level Design Constraints}}.
\newblock \bibinfo{publisher}{Association for Computing Machinery},
  \bibinfo{address}{New York, NY, USA}, \bibinfo{pages}{1–13}.
\newblock
\showISBNx{9781450367080}
\urldef\tempurl%
\url{https://doi.org/10.1145/3313831.3376593}
\showURL{%
\tempurl}


\bibitem[Weiss(1994)]%
        {openCoding}
\bibfield{author}{\bibinfo{person}{Robert~Stuart Weiss}.}
  \bibinfo{year}{1994}\natexlab{}.
\newblock \bibinfo{booktitle}{\emph{Learning from strangers: The art and method
  of qualitative interview studies}}.
\newblock \bibinfo{publisher}{Free Press}. ix, 246 pages.
\newblock
\showISBNx{978-0-02-934625-9}


\bibitem[Widmer and Kubat(1996)]%
        {concept_drift}
\bibfield{author}{\bibinfo{person}{Gerhard Widmer} {and}
  \bibinfo{person}{Miroslav Kubat}.} \bibinfo{year}{1996}\natexlab{}.
\newblock \showarticletitle{Learning in the Presence of Concept Drift and
  Hidden Contexts}.
\newblock \bibinfo{journal}{\emph{Mach. Learn.}} \bibinfo{volume}{23},
  \bibinfo{number}{1} (\bibinfo{date}{April} \bibinfo{year}{1996}),
  \bibinfo{pages}{69–101}.
\newblock
\showISSN{0885-6125}
\urldef\tempurl%
\url{https://doi.org/10.1023/A:1018046501280}
\showDOI{\tempurl}


\bibitem[Wu et~al\mbox{.}(2021)]%
        {wu_screen_parsing}
\bibfield{author}{\bibinfo{person}{Jason Wu}, \bibinfo{person}{Xiaoyi Zhang},
  \bibinfo{person}{Jeffrey Nichols}, {and} \bibinfo{person}{Jeffrey~P.
  Bigham}.} \bibinfo{year}{2021}\natexlab{}.
\newblock \showarticletitle{Screen Parsing: Towards Reverse Engineering of {UI}
  Models from Screenshots}.
\newblock \bibinfo{journal}{\emph{CoRR}}  \bibinfo{volume}{abs/2109.08763}
  (\bibinfo{year}{2021}).
\newblock
\showeprint[arXiv]{2109.08763}
\urldef\tempurl%
\url{https://arxiv.org/abs/2109.08763}
\showURL{%
\tempurl}


\bibitem[Wu et~al\mbox{.}(2020)]%
        {wu_animations_2020}
\bibfield{author}{\bibinfo{person}{Ziming Wu}, \bibinfo{person}{Yulun Jiang},
  \bibinfo{person}{Yiding Liu}, {and} \bibinfo{person}{Xiaojuan Ma}.}
  \bibinfo{year}{2020}\natexlab{}.
\newblock \showarticletitle{Predicting and {Diagnosing} {User} {Engagement}
  with {Mobile} {UI} {Animation} via a {Data}-{Driven} {Approach}}. In
  \bibinfo{booktitle}{\emph{Proceedings of the 2020 {CHI} {Conference} on
  {Human} {Factors} in {Computing} {Systems}}}. \bibinfo{publisher}{ACM},
  \bibinfo{address}{Honolulu HI USA}, \bibinfo{pages}{1--13}.
\newblock
\showISBNx{978-1-4503-6708-0}
\urldef\tempurl%
\url{https://doi.org/10.1145/3313831.3376324}
\showDOI{\tempurl}


\bibitem[Yang et~al\mbox{.}(2018)]%
        {yang_uxml}
\bibfield{author}{\bibinfo{person}{Qian Yang}, \bibinfo{person}{Alex Scuito},
  \bibinfo{person}{John Zimmerman}, \bibinfo{person}{Jodi Forlizzi}, {and}
  \bibinfo{person}{Aaron Steinfeld}.} \bibinfo{year}{2018}\natexlab{}.
\newblock \showarticletitle{Investigating How Experienced UX Designers
  Effectively Work with Machine Learning}. In
  \bibinfo{booktitle}{\emph{Proceedings of the 2018 Designing Interactive
  Systems Conference}} (Hong Kong, China) \emph{(\bibinfo{series}{DIS '18})}.
  \bibinfo{publisher}{Association for Computing Machinery},
  \bibinfo{address}{New York, NY, USA}, \bibinfo{pages}{585–596}.
\newblock
\showISBNx{9781450351980}
\urldef\tempurl%
\url{https://doi.org/10.1145/3196709.3196730}
\showDOI{\tempurl}


\bibitem[Zang et~al\mbox{.}(2021)]%
        {zang_icons}
\bibfield{author}{\bibinfo{person}{Xiaoxue Zang}, \bibinfo{person}{Ying Xu},
  {and} \bibinfo{person}{Jindong Chen}.} \bibinfo{year}{2021}\natexlab{}.
\newblock \bibinfo{booktitle}{\emph{Multimodal Icon Annotation For Mobile
  Applications}}.
\newblock \bibinfo{publisher}{Association for Computing Machinery},
  \bibinfo{address}{New York, NY, USA}.
\newblock
\showISBNx{9781450383288}
\urldef\tempurl%
\url{https://doi.org/10.1145/3447526.3472064}
\showURL{%
\tempurl}


\bibitem[Zhang et~al\mbox{.}(2021)]%
        {screen_recognition_apple}
\bibfield{author}{\bibinfo{person}{Xiaoyi Zhang}, \bibinfo{person}{Lilian de
  Greef}, \bibinfo{person}{Amanda Swearngin}, \bibinfo{person}{Samuel White},
  \bibinfo{person}{Kyle Murray}, \bibinfo{person}{Lisa Yu}, \bibinfo{person}{Qi
  Shan}, \bibinfo{person}{Jeffrey Nichols}, \bibinfo{person}{Jason Wu},
  \bibinfo{person}{Chris Fleizach}, \bibinfo{person}{Aaron Everitt}, {and}
  \bibinfo{person}{Jeffrey~P Bigham}.} \bibinfo{year}{2021}\natexlab{}.
\newblock \bibinfo{booktitle}{\emph{Screen Recognition: Creating Accessibility
  Metadata for Mobile Applications from Pixels}}.
\newblock \bibinfo{publisher}{Association for Computing Machinery},
  \bibinfo{address}{New York, NY, USA}.
\newblock
\showISBNx{9781450380966}
\urldef\tempurl%
\url{https://doi.org/10.1145/3411764.3445186}
\showURL{%
\tempurl}


\end{thebibliography}
